\documentclass[aps,reprint,superscriptaddress]{revtex4-2}

\usepackage[utf8]{inputenc}
\usepackage{amsmath, amssymb}

\usepackage{graphicx}
\graphicspath{{./figs/}}
\usepackage[textwidth=7.09in,top=1in,bottom=1in]{geometry}

\usepackage[unicode]{hyperref}
\hypersetup{
	colorlinks=true,
	citecolor=blue,
	linkcolor=black,
	urlcolor=blue,
}

\makeatletter
\renewcommand\subsection{\@startsection{subsection}{2}{\z@}%
  {-1.5ex}{0.3ex}{\noindent\normalfont\bfseries}}
\makeatother

\makeatletter
\providecommand{\SIlabel}[2]{\@namedef{SI@#1}{#2}}
\providecommand{\SIref}[1]{\@ifundefined{SI@#1}{\textbf{??}}{\@nameuse{SI@#1}}}
\makeatother

\SIlabel{fig:figS_fridge}{S1}
\SIlabel{fig:figS_2d_1d}{S2}
\SIlabel{fig:figS_steps}{S3}
\SIlabel{fig:figS_cuts_at_2G0}{S4}
\SIlabel{fig:figS_cuts_gap}{S5}
\SIlabel{fig:figS_cuts_gap_log_scale}{S6}
\SIlabel{fig:figS_beenakker_3devices}{S7}
\SIlabel{fig:figS_beenakker_vg_vmeas}{S8}
\SIlabel{fig:figS_bulb}{S9}
\SIlabel{fig:figS_magnetic_field}{S10}
\SIlabel{fig:figS_cuts_in_magnetic_field}{S11}
\SIlabel{fig:figS_pinchoffs}{S12}
\SIlabel{fig:figS_EDS}{S13}
\SIlabel{fig:figS_como_jjs}{S14}
\SIlabel{fig:figS_SEM_JJs}{S15}
\SIlabel{fig:figS_fab_steps}{S16}

\newcommand{\suppfig}[1]{Fig.~\SIref{fig:#1}}

\begin{document}
\title{Superconducting edge contact to a buried Ge quantum well}
\author{Dina Sokolova}
\affiliation{Institute of Science and Technology Austria, Am Campus 1, Klosterneuburg, 3400, Austria}
\author{Davide Costa}
\affiliation{QuTech and Kavli Institute of Nanoscience, Delft University of Technology, 2600 GA Delft, The Netherlands}
\author{Davide Degli Esposti}
\affiliation{QuTech and Kavli Institute of Nanoscience, Delft University of Technology, 2600 GA Delft, The Netherlands}
\author{Tommaso Costanzo}
\affiliation{Institute of Science and Technology Austria, Am Campus 1, Klosterneuburg, 3400, Austria}
\author{Giordano Scappucci}
\affiliation{QuTech and Kavli Institute of Nanoscience, Delft University of Technology, 2600 GA Delft, The Netherlands}
\author{Georgios Katsaros}
\email{georgios.katsaros@ist.ac.at}
\affiliation{Institute of Science and Technology Austria, Am Campus 1, Klosterneuburg, 3400, Austria}
\author{Anton V. Bubis}
\email{anton.bubis@ist.ac.at}
\affiliation{Institute of Science and Technology Austria, Am Campus 1, Klosterneuburg, 3400, Austria}

\newcommand{\reffig}[2][]{Fig.~\textbf{\ref{#2}{#1}}}
\newcommand{\refeq}[1]{Eq.~(\ref{#1})}

\newcommand{\kB}{k_\mathrm{B}}
\newcommand{\Ic}{I_\mathrm{c}}
\newcommand{\Isw}{I_\mathrm{sw}}
\newcommand{\Iexc}{I_\mathrm{exc}}
\newcommand{\Rn}{R_\mathrm{n}}
\newcommand{\Vg}{V_\mathrm{g}}
\newcommand{\vF}{v_\mathrm{F}}
\newcommand{\Ej}{E_\mathrm{J}}
\newcommand{\dvdi}{\mathrm{d}V/\mathrm{d}I}
\newcommand{\Tint}{\mathcal{T}}
\newcommand{\SM}{Supplementary Materials}

\begin{abstract}
Semiconductor--superconductor hybrid systems form the basis for a broad range of quantum phenomena, such as Andreev bound states, topological superconductivity, and hybrid quantum Hall states.
The performance of these hybrid systems can strongly benefit from the exceptional electronic properties of heterostructures with deeply buried quantum wells.
However, establishing high-quality superconducting contacts to such quantum wells remains challenging.
Here, we address this problem with an edge-contact approach that combines anisotropic dry etching to access the quantum well followed by wet etching for interface preparation.
Josephson junctions fabricated using this approach consistently exhibit near-unity interface transparency, while quantum point contacts show strong Andreev enhancement, with the first conductance plateau approaching $4e^2/h$, the limit of perfect Andreev reflection.
In superconducting quantum point contacts, the high contact quality is further manifested by quantized supercurrent.
These results establish the combination of dry and wet etching as a robust strategy for integrating deeply buried quantum wells into superconducting hybrid devices.
\end{abstract}

\maketitle

\section*{Main}
Hybrid semiconductor--superconductor devices use the proximity effect~\cite{deutscher_degennes1969} to introduce superconducting correlations into gate-tunable electronic systems.
In semiconductor weak links, carrier density and transmission set the Josephson coupling and current--phase relation~\cite{golubov2004}, making the junction an adjustable superconducting circuit element~\cite{pitavidal2025}.
The Andreev bound states that mediate this coupling can be isolated in a proximitized quantum dot, a strongly confined semiconducting region tunnel coupled to a superconductor~\cite{deacon2010, pillet2010, lee2013, lakic2025, fabris2026}.
Coupling two quantum dots to a common superconductor enables crossed Andreev reflection, which splits Cooper pairs into spatially separated electrons~\cite{hofstetter2009, herrmann2010, bordoloi2022, wang2022}.
Balanced against elastic cotunneling, the same process provides the ingredients for minimal Kitaev chains~\cite{leijnse2012, dvir2023, tenhaaf2024}.
Scaling coupled-dot architectures to larger arrays requires low-defect-density semiconductors with reproducible tunnel and proximity couplings across many sites~\cite{sau2012, bordin2025}.
Beyond confined systems, numerous proposals seek to couple superconductivity to the integer quantum Hall edge states~\cite{ma1993, vanostaay2011, finocchiaro2018, khrapai2023, parfenov2026}.
Despite evidence for chiral supercurrents in graphene~\cite{vignaud2023}, experiments in semiconductor quantum wells (QWs) have so far been limited to Andreev processes at quantum Hall edges~\cite{wan2015, hatefipour2022, akiho2024}, and coherent coupling to fractional edges remains an outstanding challenge~\cite{clarke2013, guel2022}.
Across this range of hybrid devices, a central material requirement is to combine a low-disorder semiconductor with a transparent, well-controlled superconducting interface.

For two-dimensional semiconductors, disorder is often reduced by moving the electronic system away from the surface.
Thick barriers separate carriers from surface disorder, improving mobility and percolation density, therefore QWs in the cleanest heterostructures can lie tens to hundreds of nanometers below the surface~\cite{scappucci2021mrs}.
Conventional top contacts impose the opposite constraint: strong superconducting coupling favors a quantum well close to the surface.
Existing hybrid platforms, therefore, commonly use shallow wells, contacted either through epitaxial superconductor growth~\cite{shabani2016, zhang2023} or after controlled chemical surface preparation and prompt metal deposition~\cite{heedt2021, valentini2024, fabris2026}.
Deeper quantum wells are often contacted by locally removing the barrier, either using dry etching~\cite{aggarwal2021, leblanc2024} or by wet etching followed by \textit{in situ} Ar cleaning~\cite{bendias2018, vigneau2019, ke2019}, making the superconducting interface sensitive to plasma-induced damage and contamination.
Alternatively, solid-state reaction can drive the metal towards the QW~\cite{hendrickx2019, tosato2023, lakic2025}, but it restricts the choice of superconducting material~\cite{li2026, viswanathan2026} and remains difficult to predictably extend to deep buried wells.

Here, we form superconducting contacts to a buried quantum well through its edge.
A plasma-etched trench through the heterostructure provides precise control over a deep, anisotropic etch profile and exposes the QW at its sidewall.
A subsequent gentle wet-chemical etch removes plasma residues and leaves a clean semiconductor surface prepared for direct metalization.
We demonstrate the approach in Ge/SiGe heterostructures~\cite{scappucci2021} with a quantum well 70~nm below the surface (similar growth to Ref.~\cite{stehouwer2023}), using aluminum as a prototypical superconductor.
The interface transparencies extracted from transport measurements in Josephson junctions approach unity.
Quantum point contacts defined at the superconducting edge exhibit subgap conductance steps close to twice their normal-state values, in close agreement with the prediction for a ballistic constriction with a transparent Andreev interface.
The usual trade-off between electronic quality and superconducting coupling is thus no longer dictated by the contact architecture, but can instead be independently set through heterostructure design and the choice of superconductor.

\begin{figure*}[!ht]
    \centering
    \includegraphics[width=6.20in]{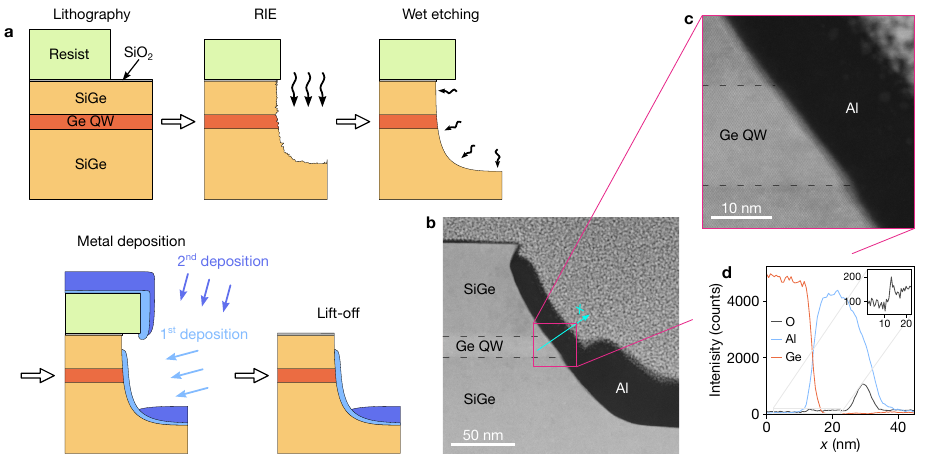}
    \caption{\textbf{Combined dry- and wet-etch fabrication process for superconducting edge contacts.}
    \textbf{a},~Cross-sectional schematics of the fabrication process.
    Starting from the pristine Ge/SiGe heterostructure (similar growth to Ref.~\cite{stehouwer2023}), dry etching defines the device geometry and exposes the Ge QW at a sidewall.
    A subsequent wet etch employed with peracetic acid~\cite{baraissov2019} removes plasma residues and potentially damaged material while creating an undercut beneath the resist.
    Following metal deposition establishes direct superconducting contact to the quantum well and provides routing to the bondpad.
    \textbf{b,\,c},~HAADF-STEM images of an Al edge contact to the Ge QW after lift-off.
    Boundaries of the Ge QW with SiGe spacers are indicated with dashed lines.
    \textbf{d},~EDS line scan along the cyan line (panel~\textbf{b}) resolves a small oxygen peak at the Al/Ge interface.}
    \label{fig1}
\end{figure*}

\begin{figure*}[!ht]
    \centering
    \includegraphics[width=6.0in]{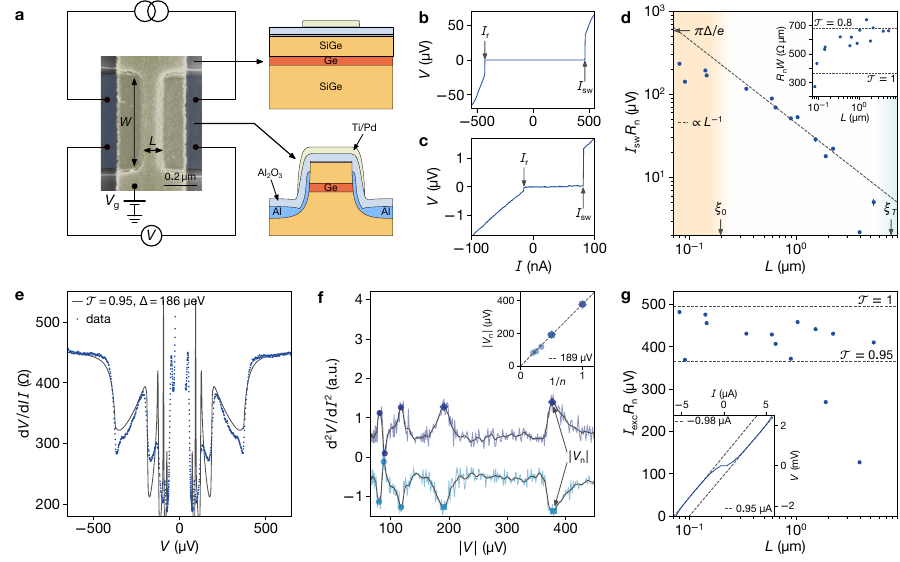}
    \caption{\textbf{Highly transparent Josephson junctions.}
    \textbf{a},~False-color SEM image of one of the shortest Josephson junctions.
    The schematics on the right show cross-sections along the current-flow direction: outside the contact region (top) and through the center of the junction (bottom).
    \textbf{b,\,c},~$IV$ curves of the shortest, $L=80$~nm (\textbf{b}), and longest, $L=5.35$~$\mu$m (\textbf{c}), junctions.
    \textbf{d},~$\Isw\Rn$ product for Josephson junctions with different geometries.
    For each junction, the mean value of $\Isw$ is calculated from 50 $IV$ curves.
    The normal-state resistance $\Rn$ was extracted from the high-bias $IV$ slope over the range $1~\mathrm{mV}<|V|<2~\mathrm{mV}$.
    Inset: Product $\Rn W$ for the same device set.
    \textbf{e},~Differential resistance $\dvdi$ as a function of the voltage drop $V$ across the shortest junction.
    A fit using the Averin--Bardas model~\cite{averinbardas1995} (see Methods) yields the effective interface transparency $\Tint=0.95$ and the superconducting gap $\Delta=186$~$\mu$eV of the Al leads.
    \textbf{f},~Second-derivative signal obtained by numerically differentiating the $\dvdi$ data in panel~\textbf{e}.
    The negative-bias branch, light blue (the positive-bias branch, dark blue), is shifted downward (upward) for clarity.
    In short, transparent junctions, low-order MAR features appear as extrema in this derivative signal.
    The voltages $V_n$ of these extrema are extracted from the smoothed traces (black) and plotted as $|V_n|$ versus $1/n$ in the inset.
    Only the $n=1,2$ features (squares), whose positions are least sensitive to $\Tint$, are included in the linear fit constrained to pass through the origin.
    The fit yields $\Delta=189$~$\mu$eV.
    \textbf{g},~Excess-current product $\Iexc\Rn$ extracted from high-bias measurements of the devices shown in panel~\textbf{d}.
    Horizontal lines indicate values predicted by the OTBK model~\cite{octavio1983, flensberg1988, mortensen1999} for selected interface transparencies $\Tint$.
    Inset: High-bias $IV$ curve of the $L=80$~nm junction, showing excess current symmetric in bias.
    All data are taken at full accumulation, where $\Rn$ no longer changes with gate voltage.}
    \label{fig2}
\end{figure*}

\begin{figure*}[!ht]
    \centering
    \includegraphics[width=7.09in]{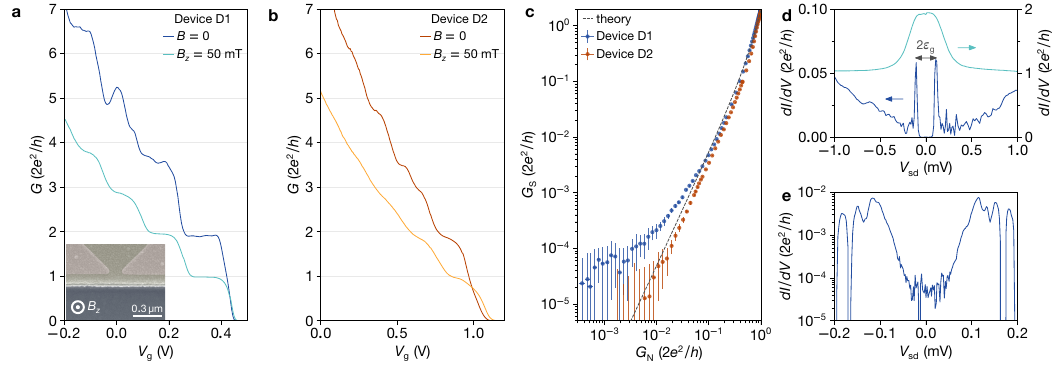}
    \caption{\textbf{Conductance doubling in QPC with superconducting edge contacts.}
    \textbf{a,\,b},~Conductance as a function of split-gate voltage, measured at zero magnetic field (darker line) and at out-of-plane field $B_z=50$~mT (lighter line), when superconductivity is suppressed.
    Panel~\textbf{a}~(\textbf{b}) shows data from a device D1 (D2) with split-gate gap of $150$~nm ($220$~nm) and gate-to-superconductor distance of $50$~nm ($0$~nm).
    Inset: false-color SEM of a device nominally identical to device D1.
    \textbf{c},~Conductance in superconducting state, $G_\mathrm{S}$, plotted vs normal-state conductance, $G_\mathrm{N}$ ($B_z=50$~mT).
    At each selected value of $\Vg$, the device was held at a fixed gate voltage while the conductance was measured at $B_z=0$ and $B_z=50$~mT, ensuring that gate-voltage hysteresis is ruled out.
    At low conductances, each point is the mean of several repeated lock-in readings and error bars denote their standard deviation.
    \textbf{d}~Differential conductance, $\mathrm{d}I/\mathrm{d}V$, of device D1, as a function of voltage bias, measured in the tunnelling regime (dark blue) and on the first conductance plateau (light blue), see full conductance map in \suppfig{figS_beenakker_vg_vmeas}.
    \textbf{e},~$\mathrm{d}I/\mathrm{d}V$ of the same device, measured deeper in the tunneling regime than in panel~\textbf{d}.}
    \label{fig3}
\end{figure*}

\subsection*{Combined dry and wet etching}
The fabrication of the superconducting contact begins by dry etching the substrate through a patterned resist mask, providing direct access to the buried quantum well, \reffig[a]{fig1}.
The edge-contact geometry relaxes the requirements for etch-depth control, while an optimized anisotropic dry etch enables the formation of nearly vertical sidewalls, preserving the lithographically defined contact geometry.
Dry etching can, however, damage semiconductor surfaces and degrade their electronic properties, as reported, for example, for HgTe quantum wells~\cite{bendias2018}.
Moreover, anisotropic plasma etching commonly relies on nonvolatile products that form a passivation layer on the sidewalls.
We therefore apply a gentle wet etch~\cite{baraissov2019} to remove plasma residues and the potentially damaged surface layer, preparing the sidewall for metal deposition.
Because only a few tens of nanometres of material are removed, the wet-etching step largely preserves the geometry defined during the preceding dry etch.
Additionally, the wet etch creates a controlled undercut beneath the resist mask, facilitating reliable lift-off even for metal deposited at grazing incidence.
After wet etching, the sample is promptly loaded into the evaporator, and a metal film is deposited in two steps to establish a superconducting contact to the buried quantum well and provide on-chip routing (for details see Methods and \SM).
Importantly, the dry etch, wet etch, and metal deposition are all performed using the same resist mask, reducing the number of lithography steps and ensuring self-alignment between the etched sidewall and the superconducting contact~\cite{liu2025}.
We use Al as a prototypical superconductor because it is among the most extensively studied materials in superconductor--semiconductor hybrids, allowing direct comparison of the resulting interface quality with previous studies~\cite{kjaergaard2016, shabani2016, hendrickx2019, heedt2021, zhang2023, akiho2024, valentini2024}.

Cross-sectional high-angle annular dark-field scanning transmission electron microscopy (HAADF-STEM) images of the Al edge contact to the Ge quantum well are shown in \reffig[b,\,c]{fig1}.
The crystalline structure of the Ge quantum well remains intact near the contact, \reffig[c]{fig1}, indicating that the combined dry- and wet-etching sequence introduces no observable structural damage.
Energy-dispersive X-ray spectroscopy (EDS) reveals an abrupt Al/Ge interface with only a weak oxygen peak (see also \suppfig{figS_EDS}).
The integrated oxygen signal at this interface is a factor of 20 lower compared to the air-exposed Al surface, where a native oxide forms after fabrication, demonstrating the effectiveness of our \textit{ex situ} interface preparation.

\subsection*{Highly transparent Josephson junctions}
Josephson junction~(JJ) measurements provide a useful benchmark for assessing the quality of edge contacts and device-to-device reproducibility.
\reffig[a]{fig2} shows a false-color scanning electron microscopy~(SEM) image of a representative device, in which two superconducting edge contacts (highlighted in blue) are coupled through the buried Ge quantum well to form a gate-tunable Josephson junction.
The device employs a top gate covering the entire mesa to accumulate holes in the quantum well and tune their density in the junction by varying the gate voltage $\Vg$.
We also investigated an alternative geometry in which the accumulation gate is confined to the region between the superconducting contacts, providing more localized electrostatic control of the weak link (see \suppfig{figS_SEM_JJs}).
No statistical difference between these geometries was observed.

To characterize device performance and fabrication reproducibility, we measured a large set of Josephson junctions with systematically varied lengths, $L$, and widths, $W$ (see Table~S1). 
The reliability of the wet-etching process enables consistent fabrication of junctions as short as $L=80$~nm (\reffig[a,\,b]{fig2}), providing access to the short-junction regime.
At full accumulation, the elastic mean free path of the two-dimensional hole gas, $\ell\approx13$~$\mu$m, exceeds all junction lengths studied, allowing the ballistic regime to be explored over nearly two decades in $L$.
We adjusted $W$ as a function of $L$ to maintain a sufficiently large Josephson energy, $\Ej=\hbar\Ic/2e$, across the device set.
The Josephson energy spans $\Ej/\kB\approx2{-}11$~K and remains well above the measurement temperature of 10~mK, allowing switching current to be clearly resolved even in the longest junctions, see \reffig[c]{fig2}.

We examine the length dependence of the $\Isw\Rn$ product across the complete device set, \reffig[d]{fig2}.
Because the junction widths vary, and $\Rn$ scales inversely with the number of transverse modes, this product largely removes the geometric width dependence.
For the shortest junctions, $\Isw \Rn$ approaches an $L$-independent plateau, consistent with a crossover to the short-junction regime characterized by $\xi_0=\hbar\vF/\pi\Delta\approx200$~nm, where $\vF$ is the Fermi velocity in Ge QW and $\Delta$ is the order parameter of Al leads~\cite{kupriyanov1981, golubov2004}.
The maximum measured value, $\Isw\Rn=230$~$\mu$V, is approximately $40\%$ of the expected zero-temperature value $\Ic\Rn=\pi\Delta/e$ for a fully transparent short junction and is comparable to the highest values reported for Al/Ge junctions~\cite{xiang2006, ridderbos2018, valentini2024}.
At larger $L$, $\Isw\Rn$ decreases approximately as $\propto L^{-1}$ over more than one decade in junction length, consistent with the long ballistic limit~\cite{kupriyanov1981, golubov2004}.
The longest junctions fall below this trend, possibly reflecting a departure from ballistic transport as $L$ approaches $\ell$.
Also, as $L$ approaches the thermal coherence length $\xi_T=\hbar v_F/(2\pi k_B T)\approx 8~\mu\mathrm{m}$, finite-temperature effects lead to a crossover from $1/L$ to exponential length dependence~\cite{kupriyanov1981,golubov2004}.
Furthermore, thermally activated escape and phase diffusion of the Josephson phase during the finite-rate current ramp make $\Isw$ a lower bound on $\Ic$, with a suppression that grows as $\Ej$ decreases.

The length dependence places the $L=80$~nm device well within the short-junction regime, allowing its finite-bias response to be analyzed quantitatively.
Under voltage bias, multiple Andreev reflections (MAR) generate a subharmonic gap structure (SGS) with features at bias voltages $\pm2\Delta/en$; their positions determine the superconducting gap, while their amplitudes and line shapes depend on the junction transmission.
In the short-junction limit, $L\ll\xi_0$, $\Delta$ is the only relevant energy scale~\cite{averinbardas1995, cuevas1996}, whereas finite junction length introduces additional subgap structure through the Andreev spectrum~\cite{ingerman2001}.
A fit of $\dvdi(V)$ to the short-junction MAR theory (see Methods), shown in \reffig[e]{fig2}, yields an effective interface transparency $\Tint=0.95$ and $\Delta=186~\mu$eV.
The positions of the SGS features provide a separate check of $\Delta$ that does not rely on fitting the full line shape.
The lowest MAR orders are particularly useful because they are least sensitive to finite-length effects and occur at the largest $|V_n|$, providing the most reliable constraint on $\Delta$.
In short and transparent junctions, these features appear only as broad kinks in $\dvdi$, making their positions difficult to determine directly.
Numerical differentiation of $\dvdi$ converts these kinks into pronounced extrema at $V_n$, as shown in \reffig[f]{fig2}.
Linear regression through the origin of $|V_n|$ versus $1/n$ for $n=1,2$ yields $\Delta=189~\mu$eV, in close agreement with the MAR fit.
We use the MAR-fit value $\Delta=186~\mu$eV for all gap-dependent quantities throughout this work.

Beyond the shortest junction, we assess contact transparency across the device set using the excess current, $\Iexc$, extracted from high-bias transport.
The $\Iexc\Rn$ values show no systematic dependence on $L$ and, within the OTBK model~\cite{octavio1983, flensberg1988, mortensen1999}, correspond to effective interface transparencies $\Tint=0.95{-}1$ for most junctions, \reffig[g]{fig2}.
The product $\Rn W$ is likewise nearly independent of $L$; comparison with a phase-averaged Landauer model for a ballistic two-dimensional weak link with two identical interfaces (see Methods) gives effective interface transparencies $\Tint=0.8{-}1$ for most devices (inset of \reffig[d]{fig2}).
The shortest junctions fall below the nominal ballistic limit, likely reflecting a mode count larger than inferred from the junction width (see \SM).
Taken together, these results show that edge-contact fabrication yields highly transmissive, reproducible interfaces while preserving ballistic transport in the buried QW.

\subsection*{Conductance doubling in quantum point contacts}
Quantum point contacts (QPCs) complement the Josephson junction measurements by resolving transport through individual modes.
Finite-bias measurements additionally reveal the subgap structure of the proximitized region. 
We therefore fabricated a series of QPC devices, with the constriction placed close to the superconducting contact.
Compared to JJs, QPC devices have a double layer of gates: the top gate, which globally accumulates holes in the quantum well, and the split gates, which define a constriction, see the inset of \reffig[a]{fig3} (see also \suppfig{figS_fab_steps}).
Beyond defining the QPC, such a two-layer gate stack is the basic building block of more complex gate-defined hybrid devices~\cite{tenhaaf2024, lakic2025}.

We characterize transport through the QPCs by measuring their conductance, $G$, as a function of split-gate voltage, $\Vg$.
For each device, the conductance was measured in the superconducting state and after suppressing superconductivity with a magnetic field; data from two devices are shown in \reffig[a,\,b]{fig3} (see also \suppfig{figS_steps}).
At the highest split-gate voltages, the constriction is fully pinched off, whereas decreasing $\Vg$ first increases the transmission of the lowest mode and then opens additional modes successively.
In the normal state, this successive opening of modes produces conductance plateaus near integer multiples of $G_0=2e^2/h$, demonstrating ballistic transport through the QPC.
Within the Beenakker model~\cite{beenakker1992}, the superconducting-state conductance is:

\begin{equation}
G_{\mathrm{S}}=2G_0\sum_n\frac{T_n^2}{(2-T_n)^2},
\label{eq:beenakker}
\end{equation}

where $T_n$ is the normal-state transmission eigenvalue of QPC mode $n$.
This relation predicts that superconducting-state conductance steps double in height, producing plateaus at integer multiples of $2G_0$.
Experimentally, strong conductance enhancement emerges with the first highly transmitting mode and persists into the multimode regime.
In the single-mode regime, where the conductance plateau is most clearly resolved, $G$ exceeds $1.7G_0$ in most devices (see \suppfig{figS_steps}).
In device D1, for example, the first superconducting-state plateau reaches approximately $1.9G_0$, corresponding to $T_1=0.99$ according to Eq.~(\ref{eq:beenakker}).

In the single-channel regime, the measured relation between $G_{\mathrm{S}}$ and $G_{\mathrm{N}}$ follows well the Beenakker prediction, with some device-to-device variation (\reffig[c]{fig3}, \suppfig{figS_beenakker_3devices} and \suppfig{figS_beenakker_vg_vmeas}).
Device D2, for example, follows the prediction throughout the experimentally accessible conductance range, whereas for device D1 $G_{\mathrm{S}}$ exceeds the predicted value for $G_{\mathrm{N}}\lesssim0.1G_0$.
This excess is consistent with a finite quasiparticle density of states (DOS) at the Fermi level in the proximitized Ge.
Within a phenomenological model that includes such a residual DOS, the deviation corresponds to approximately $2\%$ of the normal-state DOS (see \suppfig{figS_beenakker_3devices}).
The excess conductance is unlikely to result from thermal or lock-in excitation broadening because the associated energy scales are both much smaller than the induced gap.

We further characterize the proximitized region through bias-dependent measurements of the differential conductance, $\mathrm{d}I/\mathrm{d}V$.
With the QPC tuned to the first conductance plateau, the zero-bias differential conductance is approximately twice its high-bias value; the shallow central dip is consistent with a mode transmission slightly below unity, \reffig[d]{fig3}.
Similar behavior is observed in many devices (see \suppfig{figS_cuts_at_2G0}).
In the tunnelling regime, $\mathrm{d}I/\mathrm{d}V$ probes the density of states (DOS) of the proximitized region, \reffig[d]{fig3}.
The spectra display pronounced gap-edge peaks together with a suppressed but finite subgap conductance, \reffig[e]{fig3} (see also \suppfig{figS_cuts_gap_log_scale}).
For a ballistic region the expected line shape is not universal: depending on the confinement geometry and the interface transparency, the subgap spectrum can range from gapless and quasi-linear to hard and escape-rate-limited~\cite{saint-james1964, melsen1996, mcmillan1968}.
A broadened BCS DOS with plausible parameters is shown for reference in \suppfig{figS_cuts_gap_log_scale}; it reproduces the subgap conductance reasonably well but does not account for the structure above the induced gap.
We therefore adopt the positions of the gap-edge peaks as an operational measure of the induced gap, $\varepsilon_\mathrm{g}\approx 110~\mu\mathrm{eV}$, with similar values for most devices (see \suppfig{figS_cuts_gap}).

\begin{figure}[!ht]
    \centering
    \includegraphics[width=3.46in]{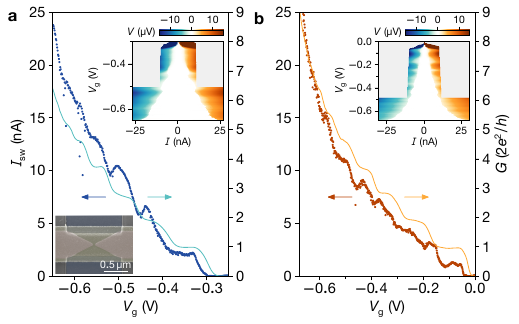}
    \caption{\textbf{Quantization of conductance and switching current in superconducting QPCs.}
    \textbf{a},~Switching current, $\Isw$ (points, left axis), and normal-state conductance, $G$ (solid line, right axis), as a function of split-gate voltage for a device with an $80$~nm split-gate separation.
    The normal-state conductance was measured at $B_{z}=50$~mT.
    Bottom inset: false-color SEM micrograph of a device nominally identical to the measured one.
    Top inset: color map of the voltage drop, $V$, as a function of current and split-gate voltage.
    The switching current was extracted using the threshold criterion $V\geq0.25~\mu\mathrm{V}$.
    \textbf{b},~Similar data for a device with a $150$~nm split-gate separation.}
    \label{fig4}
\end{figure}

\begin{figure*}[!ht]
    \centering
    \includegraphics[width=7.09in]{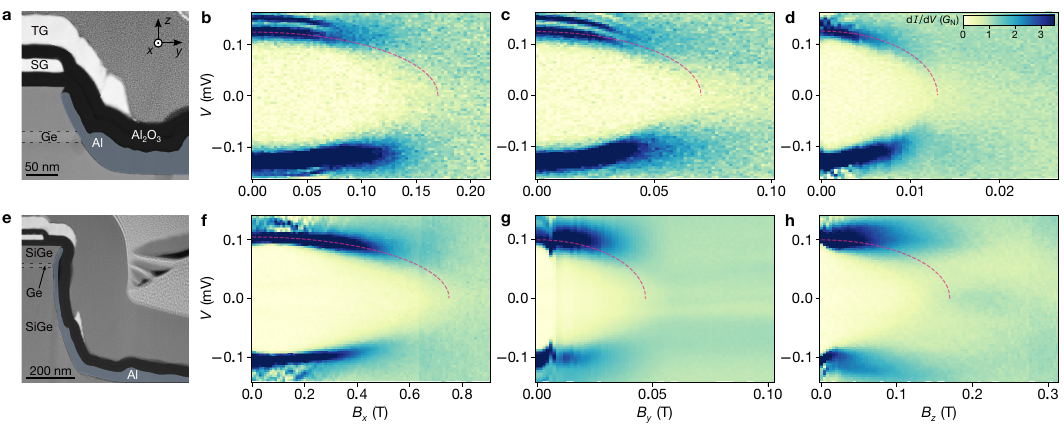}
    \caption{\textbf{Effect of the contact geometry on the magnetic-field resilience of superconducting edge contacts.}
    \textbf{a},~HAADF-STEM of one of the QPC devices studied in this work.
    Split-gate and top-gate layers are labeled SG and TG, respectively.
    Gates are separated from the substrate and each other by the atomic-layer-deposited Al$_2$O$_3$ layers.
    Al edge contact is highlighted in blue.
    \textbf{b-d},~Differential conductance in the tunneling regime as a function of bias voltage and magnetic field measured for different field directions.
    \textbf{e},~HAADF-STEM of the deeply etched device.
    \textbf{f-h},~Differential conductance measured for different field directions for a deeply etched device.
    Differential conductance for all panels is normalized by the high-field value, $G_\mathrm{N}$. 
    The dashed line is a guide to the eye based on the phenomenological dependence $\varepsilon_\mathrm{g}(B) \propto \sqrt{1-(B/B_\mathrm{c})^n}$ with $n=2$; note that the complex contact geometry may result in a different effective exponent.
    Similar measurements for two more devices presented in a~\suppfig{figS_magnetic_field}. 
    }
    \label{fig5}
\end{figure*}

\subsection*{Supercurrent quantization}
Another hallmark of ballistic transport in superconductor--semiconductor hybrids is critical-current quantization in superconducting quantum point contacts (SQPCs)~\cite{beenakker1991}.
To investigate this regime, we fabricated Josephson junctions with QPC weak links defined by different split-gate geometries.
A representative scanning electron micrograph is shown in the inset of \reffig[a]{fig4}; the split gates define the constriction, while the global top gate controls the carrier density in the quantum well.
Here, we focus on two devices with $L=370$~nm and split-gate separations of $80$~nm and $150$~nm.

With superconductivity suppressed by a magnetic field, the conductance of both devices displays well-defined plateaus close to integer multiples of $G_0$, indicating ballistic transport through discrete transverse QPC modes, \reffig{fig4}.
At zero field, both devices exhibit a gate-tunable Josephson effect (insets of \reffig[a,\,b]{fig4}).
The device with an $80$~nm split-gate separation, \reffig[a]{fig4}, shows particularly pronounced switching-current plateaus with a characteristic step height of approximately $2.5$~nA.
The transitions between these plateaus occur at the same values of $\Vg$ as the opening of successive QPC modes inferred from the normal-state conductance.
The device with a $150$~nm split-gate separation displays weaker, but still distinct, switching-current quantization with a characteristic step height of approximately $1.3$~nA (see~\suppfig{figS_bulb}).

Both step heights fall well below $e\Delta/\hbar\approx45$~nA, as in essentially all hybrid SQPCs~\cite{takayanagi1995,bauch2005,abay2013,irie2014,hendrickx2019,mikheev2021}.
The usual culprits are imperfect interface transparency and finite junction length~\cite{furusaki1992,mikheev2021}.
The interface can be excluded here: the excess current at full accumulation gives an OTBK transparency close to unity (see Table~S1), and in the quantized regime the conductance plateaus sit within a few percent of integer multiples of $G_0$ without any series-resistance subtraction.
In contrast, finite junction length can reduce $\Ic$~\cite{furusaki1992} and thus contribute to the suppressed step height.
Although $L/\xi_0\approx1.6$ at full accumulation, the split gates deplete the weak link once only a few modes remain open, and since $\xi_0\propto\vF$, the effective $L/\xi_0$ in the quantized regime can be substantially larger.
Additionally, the measured switching current likely underestimates the critical current.
If $\Ic$ were as small as the first step of $\Isw$, the corresponding Josephson energy, $\hbar\Isw/2e\approx\kB\times60$~mK, would be comparable to the electron temperature.
For such weak coupling, thermal phase diffusion would produce a sizable resistance along the entire supercurrent branch, a significant fraction of $\Rn$ assuming overdamped dynamics~\cite{ambegaokar1969}.
In contrast, the voltage remains below our noise floor of $\sim50$~nV at low bias and develops only close to $\Isw$.
This indicates a Josephson energy above the thermal energy, suggesting that $\Ic$ exceeds $\Isw$ and that the junction switches prematurely due to fluctuations.
Thus, finite junction length and premature switching together may account for a part of the suppression of the supercurrent steps.

\subsection*{Magnetic field resilience enhancement}
The edge-contact geometry provides an additional degree of freedom for engineering magnetic-field resilience.
In principle, the contact profile can be designed so that the superconducting film at the quantum-well edge is nearly parallel to a field applied perpendicular to the quantum well.
However, even with a perfectly vertical sidewall profile, the edge contact remains part of a continuous, nonplanar Al layer that provides the routing, so its magnetic response is not determined by the local film orientation alone.
We investigate these geometric effects by measuring the differential conductance of QPC devices as a function of magnetic field for three orthogonal field orientations.

\reffig[a]{fig5} shows a representative cross-sectional HAADF-STEM image of a device fabricated using our standard edge-contact process, with an etch depth of approximately $150$~nm.
In this geometry, the sidewall film is not parallel to $B_z$ and is continuously connected to the thicker Al routing layer. 
The proximity-induced conductance features disappear at only $B_z\approx13$~mT, \reffig[d]{fig5}, but persist to substantially higher fields for in-plane orientations, \reffig[b,\,c]{fig5}.
The greatest resilience is observed for $B_x$, which lies in the chip plane and is perpendicular to transport.
We attribute this anisotropy to the combined effects of the local film orientation and inverse-proximity coupling to the routing layer.
For $B_z$, the field is perpendicular to the routing layer and suppresses its superconductivity at a relatively low field; once driven normal, the routing layer suppresses superconductivity in the adjacent edge contact through the inverse proximity effect.
For an in-plane field parallel to transport, the routing layer remains superconducting and supports the edge contact up to higher fields.
Finally, $B_x$ lies in the plane of the continuous Al film both at the edge contact and along the routing path, accounting for the largest observed field resilience.

This analysis identifies the contact profile and its distance from the routing layer as design parameters, motivating the fabrication of proof-of-concept devices with a deeper, anisotropic etch.
The resulting etch depth is approximately $600$~nm, substantially increasing the distance between the quantum-well contact and the routing layer.
A representative transmission electron micrograph of a deep-contact device is shown in \reffig[e]{fig5}.
In the deep-contact devices, the critical field for $B_z$ increases by approximately a factor of ten ($B_{\mathrm{c},z}\approx 170$~mT), consistent with closer alignment of the sidewall film with $B_z$ and reduced inverse-proximity coupling to the routing layer.
A similarly pronounced improvement is observed for $B_x$, which lies in the plane of the Al layer throughout the device; the critical field reaches approximately $750$~mT (see~\suppfig{figS_cuts_in_magnetic_field}), consistent with the high in-plane critical field of thin-film Al.
In contrast, the critical field for the orthogonal in-plane direction decreases slightly, consistent with weaker proximity support from the routing layer in the deeper geometry.
Together, these results show that the etch profile controls the magnetic-field resilience and anisotropy of superconducting edge contacts through both local film orientation and inverse-proximity coupling to the routing layer.

\subsection*{Conclusions}
Electronic quality and superconducting coupling need not be traded against one another in two-dimensional hybrid devices.
Superconducting edge contacts provide direct access to a buried electronic system, allowing the QW depth to be set by heterostructure design rather than contact geometry.
Here, we demonstrate this capability in a Ge/SiGe heterostructure with a QW buried $70$~nm below the surface, using Al as a prototypical superconductor.
MAR and excess-current analyses of Josephson junctions spanning nearly two decades in length yield effective Al/Ge interface transparencies approaching unity, while the length scaling remains characteristic of ballistic transport.
Gate-defined QPCs extend this performance to individual modes: their superconducting-state conductance approaches the doubling expected for perfect Andreev reflection, and switching-current steps in QPC weak links track successive mode openings.
Deep-etch devices further show that the magnetic response is itself a design variable, with an order-of-magnitude enhancement of the out-of-plane field scale achieved by reshaping and spatially decoupling the nonplanar Al contact.

More broadly, edge access separates the interface-formation strategy from the heterostructure design and thereby widens the choice of superconducting material.
For example, Ir-derived germanides formed by solid-state reaction are superconductors with tesla-scale upper critical fields~\cite{li2026, viswanathan2026}.
With Ir deposited directly on the QW sidewall, this reaction could proceed locally at the edge rather than requiring Ir to diffuse through the full SiGe cap.
This freedom is relevant both to larger coupled structures, where uniform proximity coupling must be maintained across many elements, and to high-field hybrid systems.
Combining deep, low-disorder Ge QWs with transparent, magnetic-field-resilient contacts removes a central materials obstacle to superconductor--quantum Hall hybrids and, ultimately, coherent coupling to fractional quantum Hall edge states.

\subsection*{Acknowledgements}
Maksim Borovkov made a major contribution to the development of the atomic layer deposition recipe used in this work.
We are grateful to Jeroen Danon, Giorgio Fabris, Paul Falthansl-Scheinecker, Vadim Khrapai, Alfredo Levy~Yeyati, Devashish Shah and Evgeny Tikhonov for careful reading and comments on the manuscript.
The ISTA research was supported by the Scientific Service Units of ISTA through resources provided by the MIBA Machine Shop, the Nanofabrication, the Electron Microscopy and the Lab Support facility; the Austrian Science Fund (FWF) [grant DOI 10.55776/PAT7682124] and the European Innovation Council via the Pathfinder grant no. 101115315 (QuKiT).
For open access purposes, the authors have applied a CC BY public copyright license to any author accepted manuscript version arising from this submission.

\subsection*{Author contributions}
A.B. and G.K. conceived the project.
D.S. and A.B. fabricated the devices and performed the transport measurements.
D.C., D.D.E. and G.S. validated and provided the Ge/SiGe heterostructures.
A.B. made lamellae for TEM analysis; T.C. and A.B. performed TEM imaging. 
A.B. and G.K. supervised the work.
D.S. and A.B. wrote the manuscript with the input of G.K.
All authors discussed the results and commented on the manuscript.

\section*{Methods}
\subsection*{Devices fabrication}
Devices were fabricated from a Ge/SiGe heterostructure with a maximum hole density of $2\times 10^{11}$~cm$^{-2}$ and corresponding mobility of $1.8 \times 10^{6}$~cm$^2$/Vs.
The epitaxial deposition protocol, starting from a Ge wafer, is similar to Ref.~\cite{stehouwer2023}.
We used electron-beam lithography for all patterning. 
Superconducting edge contacts were fabricated using reactive ion etching (RIE), wet chemical etching, and angled aluminum deposition.
Our standard contacts (Figs.~\textbf{1-5d}) were formed by etching 120--125 nm of material using an SF$_6$:O$_2$:CHF$_3$ plasma, whereas deep contacts (\reffig[e-h]{fig5}) were formed by etching approximately 600 nm using an SF$_6$:CHF$_3$ plasma. 
A subsequent wet etch of the exposed sidewalls in HF:H$_2$O$_2$:CH$_3$COOH:H$_2$O removed an additional 20--25~nm of material and was followed by an HF dip.
Samples were then transferred into a high-vacuum electron-beam evaporator.
Aluminum contact deposition involved multiple angles, starting with 15~nm of Al at $75^\circ$, followed by a $180^\circ$ sample rotation and an equivalent deposition on the opposite sidewall.
A subsequent 30~nm Al deposition at $15^\circ$ from both sides completed the contact geometry.
Mesa was defined by RIE using an SF$_6$:O$_2$:CHF$_3$ plasma. 
A 22 nm thick Al$_2$O$_3$ dielectric layer was deposited by atomic layer deposition at $80\,^\circ$C (see \SM, \suppfig{figS_EDS} and \suppfig{figS_como_jjs}).
The gates consisted of Ti/Pd stacks deposited by electron-beam evaporation.
QPC and SQPC devices additionally contained a second 22 nm thick Al$_2$O$_3$ dielectric layer and a top-gate stack consisting of Ti/Pd.

\subsection*{Measurements}
We used two dilution refrigerators for the electrical measurements, both equipped with three-axis vector magnets, with base temperatures of 10~mK (\reffig{fig2}, \reffig[f-h]{fig5}) and 70~mK (\reffig{fig3}\,--\,\reffig[d]{fig5}).

All measurements were performed in a quasi-four-probe configuration: the fridge wiring resistance is excluded, while the on-chip routing contributes in series, as in a two-probe measurement (see \suppfig{figS_fridge}).
Josephson junctions were current-biased, and the voltage drop across the junction was recorded.
For the QPC and SQPC devices, we applied a voltage bias across the series combination of the fridge lines and the device, and recorded both the current through the device and the voltage drop across it.

Differential resistance and conductance were measured using a standard lock-in technique, with a $1~\mathrm{nA}$ RMS current excitation for the current-biased measurements, \reffig[e]{fig2}, and a $5~\mu\mathrm{V}$ RMS voltage excitation for the voltage-biased measurements.

No series resistance was subtracted from the presented data. 
For QPC devices, the measured conductance includes an approximately 60~$\Omega$ series resistance originating from the device mesa.

\subsection*{Interface transparency from transport}
We extract the Al/Ge interface transparency in three ways -- from the normal-state resistance, the excess current, and the MAR line shape.
In all three analyses, each Al/Ge interface is described within the BTK model~\cite{blonder1982} as a delta-function barrier of dimensionless strength $Z$, and we report the resulting single-interface transparency at normal incidence, $\Tint=1/(1+Z^{2})$.
Fermi-velocity mismatch between Al and the Ge quantum well is not treated separately, so $\Tint$ should be understood as an effective transparency that includes any contribution from the mismatch.
This treatment does not imply that the mismatch is negligible: using bulk quantum-well parameters, mismatch alone would limit the interface transparency to approximately $30\%$, substantially below our extracted values.
The origin of the apparently weak effective barrier remains unresolved and is discussed in \SM.

A transverse mode incident at angle $\theta$ to the interface normal conserves its momentum parallel to the interface and therefore sees an effective barrier $Z/\cos\theta$~\cite{kupka1997,mortensen1999}, giving
\begin{equation}
  \Tint_{\theta}=\frac{\cos^{2}\theta}{\cos^{2}\theta+Z^{2}}
  =\frac{\Tint\cos^{2}\theta}{1-\Tint\sin^{2}\theta},
  \label{eq:Ttheta}
\end{equation}
which reduces to $\Tint$ at normal incidence and vanishes at grazing incidence.
The transverse modes of a wide junction are uniformly spaced in $\sin\theta$, so mode averages are taken with the measure $\mathrm{d}(\sin\theta)$ over $(0,\,1)$, as in the two-dimensional angular average of Ref.~\cite{mortensen1999}.

\emph{Normal-state resistance.}
Let $\tau$ denote a transmission eigenvalue of the junction.
The normal-state conductance is $1/\Rn=(2e^{2}/h\,\langle\tau\rangle)(k_{\mathrm{F}}W/\pi)$, where the first factor is the average conductance per mode and the second counts the number of modes.
Each mode traverses two identical interfaces and has a total transmission:
\begin{equation}
  \tau(\theta,\phi)=\frac{\Tint_{\theta}^{2}}
  {\bigl|1-(1-\Tint_{\theta})e^{i\phi}\bigr|^{2}},
  \label{eq:tauFP}
\end{equation}
where $\phi$ is the round-trip phase.
With $k_{\mathrm{F}}L\gg1$, we take $\phi$ uniformly distributed and  obtain the transmission $\Tint_{\theta}/(2-\Tint_{\theta})$.
In the incoherent limit, summing the probabilities of repeated reflections gives the same expression~\cite{datta1995}; thus, $\Rn$ probes only the mean transmission.
Averaging over the modes gives
\begin{align}
  \langle\tau\rangle
  &=\int_{0}^{1}\!\mathrm{d}(\sin\theta)\,
  \frac{\Tint_{\theta}}{2-\Tint_{\theta}}
  \notag\\
  &=1-\frac{2(1-\Tint)}{\sqrt{\Tint(2-\Tint)}}\,
  \operatorname{artanh}\sqrt{\frac{\Tint}{2-\Tint}},
  \label{eq:taumean}
\end{align}
and $\Rn W=(h/2e^{2})(\pi/k_{\mathrm{F}})/\langle\tau\rangle$.

\emph{Excess current.}
Most devices in the set have $L>\xi_{0}$ and are therefore outside the short-junction limit assumed by the Averin--Bardas theory.
To treat the whole set on the same footing, we evaluate $\Iexc$ within the OTBK model~\cite{octavio1983,flensberg1988}, which treats the two interfaces as spatially separated but neglects the phase accumulated between successive Andreev reflections.
The angle-resolved barriers and mode averaging are treated as above~\cite{mortensen1999}.
Note that in the unity-transmission limit, $\Tint=1$, both coherent and incoherent descriptions coincide and give $\Iexc\Rn=8/3\,(\Delta/e)$.
At reduced transparency, however, the $\Iexc\Rn$ product becomes model-dependent.

\emph{MAR line shape.}
For the shortest junction, we use the coherent MAR theory of Averin and Bardas~\cite{averinbardas1995}, which treats the constriction as independent channels with transmission eigenvalues $\tau$ and yields the current and its bias derivative for given $\tau$ and $\Delta$.
The eigenvalues are again generated from $\Tint$, but here the phase dependence of Eq.~(\ref{eq:tauFP}) is retained.
Again, we take the round-trip phase to be uniformly distributed, giving the probability density:
\begin{equation}
  P_{\Tint}(\tau)=\int_{0}^{1}\!\mathrm{d}(\sin\theta)
  \int_{0}^{2\pi}\!\frac{\mathrm{d}\phi}{2\pi}\;
  \delta\bigl[\tau-\tau(\theta,\phi)\bigr],
  \label{eq:Ptau}
\end{equation}
which contains a substantial population of near-ballistic channels because $\tau\rightarrow1$ on resonance for any $\Tint$.
Unlike $\Rn$, the subgap current is strongly nonlinear in $\tau$, so the MAR current must be averaged over $P_{\Tint}(\tau)$ rather than evaluated at $\langle\tau\rangle$.
The measured $\mathrm{d}V/\mathrm{d}I(V)$ is fitted by least squares with two free parameters, $\Tint$ and $\Delta$.
The model is normalized to its own high-bias plateau, so the mode count cancels from the fit.
We verified that a single-transparency fit, in which all channels share the same $\tau$, reproduces the low-order subharmonic structure and returns a comparable gap (see \suppfig{figS_2d_1d}).

\subsection*{Transmission electron microscopy}
Cross-sectional lamellae were prepared with a Xe plasma FIB Helios~G4 following standard lift-out protocols.
STEM measurements were performed on a JEOL JEM-2800 operated at 200~kV with a convergence semi-angle of about 10.5~mrad.
HAADF images (\reffig[b,\,c]{fig1}, \reffig[a,\,e]{fig5}, \suppfig{figS_EDS}, \suppfig{figS_como_jjs}) were acquired with a probe current of about 50~pA and an inner collection semi-angle of about 50~mrad.
ADF images (\suppfig{figS_EDS}) were recorded at a longer camera length, corresponding to an inner collection semi-angle of about 16~mrad.
For EDS mapping, the probe current was increased to about 600~pA.

\bibliography{main}
\end{document}


\title{Supplementary Materials for\\``Superconducting edge contact to a buried Ge quantum well''}
\author{Dina Sokolova}
\affiliation{Institute of Science and Technology Austria, Am Campus 1, Klosterneuburg, 3400, Austria}
\author{Davide Costa}
\affiliation{QuTech and Kavli Institute of Nanoscience, Delft University of Technology, 2600 GA Delft, The Netherlands}
\author{Davide Degli Esposti}
\affiliation{QuTech and Kavli Institute of Nanoscience, Delft University of Technology, 2600 GA Delft, The Netherlands}
\author{Tommaso Costanzo}
\affiliation{Institute of Science and Technology Austria, Am Campus 1, Klosterneuburg, 3400, Austria}
\author{Giordano Scappucci}
\affiliation{QuTech and Kavli Institute of Nanoscience, Delft University of Technology, 2600 GA Delft, The Netherlands}
\author{Georgios Katsaros}
\affiliation{Institute of Science and Technology Austria, Am Campus 1, Klosterneuburg, 3400, Austria}
\author{Anton V. Bubis}
\affiliation{Institute of Science and Technology Austria, Am Campus 1, Klosterneuburg, 3400, Austria}

\newcommand{\reffig}[2][]{Fig.~\ref{#2}{#1}}
\newcommand{\refeq}[1]{Eq.~(\ref{#1})}

\newcommand{\kB}{k_\mathrm{B}}
\newcommand{\Ic}{I_\mathrm{c}}
\newcommand{\Isw}{I_\mathrm{sw}}
\newcommand{\Iexc}{I_\mathrm{exc}}
\newcommand{\Rn}{R_\mathrm{n}}
\newcommand{\Vg}{V_\mathrm{g}}
\newcommand{\vF}{v_\mathrm{F}}
\newcommand{\Ej}{E_\mathrm{J}}
\newcommand{\dvdi}{\mathrm{d}V/\mathrm{d}I}
\newcommand{\Tint}{\mathcal{T}}

\maketitle

\section{Fabrication:}
\subsection{Wet etching}
The stock solution for wet etching is a mixture of HF\,:\,H$_2$O$_2$\,:\,CH$_3$COOH\,:\,H$_2$O~\cite{baraissov2019} in a volumetric ratio of 2\,:\,36.4\,:\,42\,:\,15.6, \#2 below.
For a few days after preparation, the etching rate usually changes slightly, so we let it stabilize for at least two weeks before usage.
Prior to wet etching of the sample, the stock solution is diluted in deionized water \#2\,:\,H$_2$O\:=\:1\,:\,15, yielding an etch rate of approximately \SI{0.7}{\nm\per\s}.
The etching rate has a strong dependence on Ge content, for wafers with a spacer composition of Si$_{0.3}$Ge$_{0.7}$ we usually use a dilution ratio of 1:7 and achieve similar etch rates.
This results in a mushroom-like cross-section for Si$_{0.3}$Ge$_{0.7}$ spacers, see \reffig[g-i]{fig:figS_como_jjs}, as lateral etching of the Ge QW is higher than that of the SiGe spacer.

The choice of resist stack was found to be critical for minimizing resist peeling and solution creeping between the resist and the substrate.
The most reliable results were obtained using an ``AR 300-80 new'' adhesion promoter in combination with a CSAR~62 (AR-P 6200.13) resist.

\subsection{Dry etching}
Most of the samples were fabricated using our group standard RIE plasma recipe SF$_6$\,:\,O$_2$\,:\,CHF$_3$ with a gas flow ratio of 9\,:\,12\,:\,3, 10\si{\milli\torr} and \SI{10}{\W} RF power (resulting in a DC bias voltage of \SIrange{52}{54}{\V}).
This plasma gives a low-anisotropy etching profile with an etching rate of about \SI{36}{\nm\per\min} (almost independent on Ge content within Si$_{0.3}$Ge$_{0.7}$ to pure Ge).

For deep etching, an SF$_6$:CHF$_3$ plasma with a gas flow ratio of 4:16 was used at a chamber pressure of \SI{3}{\milli\torr} and an RF power of \SI{60}{\W}, resulting in a DC self-bias voltage of approximately \SI{330}{\V}.
The corresponding etch rate of Si$_{0.2}$Ge$_{0.8}$ was approximately \SI{40}{\nm\per\min}, producing more vertical sidewalls compared due to high degree of anisotropy.

\subsection{Low temperature ALD}
To minimize thermally induced propagation of Al into the SiGe heterostructure, the ALD process used to deposit the gate dielectric was performed at a reduced temperature of \SI{80}{\celsius}.
Nevertheless, signatures of Al redistribution are occasionally observed even at this temperature.
\reffig[a,\,b]{fig:figS_EDS} shows HAADF-STEM micrographs from several devices in which an Al signal is detected within the quantum well and the surrounding SiGe.
Note that this effect is not specific to the superconducting contact or the quantum-well interface, for example, \reffig[c]{fig:figS_EDS} shows a routing layer in which Al propagated into the underlying SiGe.

To determine whether thermal annealing at the ALD temperature is an important ingredient for transparent contacts, we performed a control experiment on a Si$_{0.3}$Ge$_{0.7}$/Ge heterostructure with a quantum well \SI{20}{\nm} below the surface.
The QW in this heterostructure hosts holes even at zero gate voltage, allowing us to fabricate Josephson junctions within a single fabrication step.
In such single-layer test structures, the mesa-isolation step was also omitted, and as a consequence, the junctions remain electrically shorted through the surrounding quantum well, preventing a reliable extraction of the normal-state resistance.
Nevertheless, a pronounced switching current is observed at \SI{300}{\milli\K} before any thermal treatment (\reffig[a]{fig:figS_como_jjs}).

The sample was subsequently exposed to \SI{80}{\celsius} for the same duration and under the same pressure conditions as used during ALD, but without depositing the oxide.
This procedure isolates the effect of the thermal exposure from possible changes associated with dielectric deposition.
The current--voltage characteristics measured before and after thermal treatment do not show a substantial change in the switching current~\reffig[a]{fig:figS_como_jjs}.
The observation of a sizeable supercurrent before thermal treatment, together with the weak influence of the thermal exposure on $\Isw$ despite the redistribution of Al, indicates that annealing at the ALD temperature is not required to establish superconducting coupling to the quantum well.

\section{Possible modification of the carrier density near the superconducting contact}
Several experimental observations suggest that the electronic properties of the quantum well may be modified in the vicinity of the superconducting contacts.
One possible interpretation is an enhanced hole density close to the contacts, although the measurements discussed below do not provide a direct measurement of the local carrier density.
Here, we summarize the experimental observations motivating this interpretation.

\textbf{Normal state resistance.} Normal state resistance of the Josephson junctions provides a first indication that the electronic properties of the quantum well may be modified close to the superconducting contacts.
For a ballistic junction, the resistance--width product can be estimated as $R_\mathrm{n}W =\frac{h}{2e^2}\frac{\pi}{k_\mathrm{F}}\frac{1}{\langle\tau\rangle}$, where $k_\mathrm{F}$ is the Fermi wave vector and $\langle\tau\rangle$ is the average transmission of the occupied modes.
For few \si{\um}-long junctions, $R_\mathrm{n}$ is approximately independent of junction length, forming a plateau consistent with the ballistic estimate.
As the junction length is reduced, however, $R_\mathrm{n}$ progressively deviates from this plateau and decreases.
For the shortest junction, the measured resistance falls below the ballistic estimate, reaching a value lower than expected even for unity transmission.

In principle, such a reduction could result from an underestimate of the number of conducting modes, for example if the effective width of the junction were larger than its lithographic width.
However, accounting for the resistance of the shortest junctions solely in this way would require a substantial number of additional modes extending beyond the nominal junction region, for which we find no independent experimental indication.
Alternatively, an increased local carrier density would increase the number of modes and therefore reduce the ballistic resistance--width product.
Therefore, the resistance alone cannot distinguish this interpretation from geometric uncertainties in the effective number of modes.

\textbf{Pinch-off characteristics.} Second indication of a possible modification of the carrier density near the superconducting contacts is provided by the pinch-off characteristics of junctions with different lengths.
To allow a meaningful comparison between devices, we consider only the first gate sweep after cooldown, since the pinch-off voltage can shift substantially depending on the preceding gate-voltage history.
For junctions with $L>0.3$~\si{\micro\meter}, the extracted pinch-off voltages are approximately independent of junction length and cluster around the same value (\reffig{fig:figS_pinchoffs}).
In contrast, the shorter junctions show qualitatively different behavior.
The junction with $L=0.14$~\si{\um} conducts already at $\Vg=0$ and exhibits a positive pinch-off voltage.
The shortest junction with $L=0.08$~\si{\um} cannot be pinched off over the measured range of gate voltages.

Such a behavior of the shortest junctions is consistent with a modification of the electrostatic potential in the vicinity of the superconducting contacts.
If the hole density is induced over a finite distance into the junction, the contact-proximal regions would occupy an increasingly large fraction of the junction as $L$ is reduced.
For sufficiently short junctions, the regions with induced density could overlap, resulting in a conducting region at $\Vg=0$.
The pinch-off measurements therefore provide suggestive, although indirect, evidence for a contact-induced modification of the carrier density.

\textbf{Effective interface barrier.} One more indication that the electronic properties close to the superconducting contacts may differ from those of the bulk quantum well is provided by the exceptionally high interface transparency.
Within the BTK framework~\cite{blonder1983, mortensen1999}, Fermi-velocity mismatch contributes to the effective interface barrier according to $Z_{\mathrm{eff}}^2 = Z^2+\frac{(1-r)^2}{4r}$, where $Z$ is the intrinsic interface barrier and $r=v_{\mathrm{F,Ge}}/v_{\mathrm{F,Al}}$ is the ratio of Fermi velocities.
Using the bulk carrier density $n=2\times10^{11}$~\si{\per\cm\squared}, $m^*=0.075m_\mathrm{e}$~\cite{costa2026}, and $v_{\mathrm{F,Al}}=2\times10^6$~\si[per-mode=symbol]{\m\per\s}~\cite{akiho2024}, the velocity mismatch alone gives $Z_{\mathrm{eff}}^2\approx2.4$, even for $Z=0$.
The corresponding interface transparency, $\Tint=\frac{1}{1+Z_{\mathrm{eff}}^2}$, is approximately $29\%$, substantially below the transparency inferred experimentally from the excess-current analysis.

One possible explanation for this discrepancy is an increased hole density in the quantum well close to the superconducting contact.
A higher local density would reduce the Fermi-velocity mismatch and its contribution to $Z_{\mathrm{eff}}$.
We emphasize, however, that this comparison relies on a simplified BTK description of the interface and therefore provides only indirect support for a modified local carrier density.

Taken together, the normal-state resistance, pinch-off characteristics, and unexpectedly small effective interface barrier are consistent with a charge accumulation near the superconducting contact.
However, none of the data presented here constitutes a direct measurement of the spatial carrier-density profile.
Such a modification could potentially originate from Fermi-level pinning and the resulting local band alignment at the semiconductor--superconductor interface, or from strain induced by the superconducting contact or gate stack.
Metal contacts to Ge are known to pin the Fermi level close to the valence-band edge~\cite{dimoulas2006, nishimura2007}, which would favor hole accumulation near the Al contact.
A comparable metal-induced accumulation has recently been observed directly in InAs quantum wells beneath several superconducting films, where each metal creates a pinned interface subband with a density up to five times that of the well~\cite{coombs2026}.

\begin{figure*}[h]
    \centering
    \includegraphics[width=5.9in]{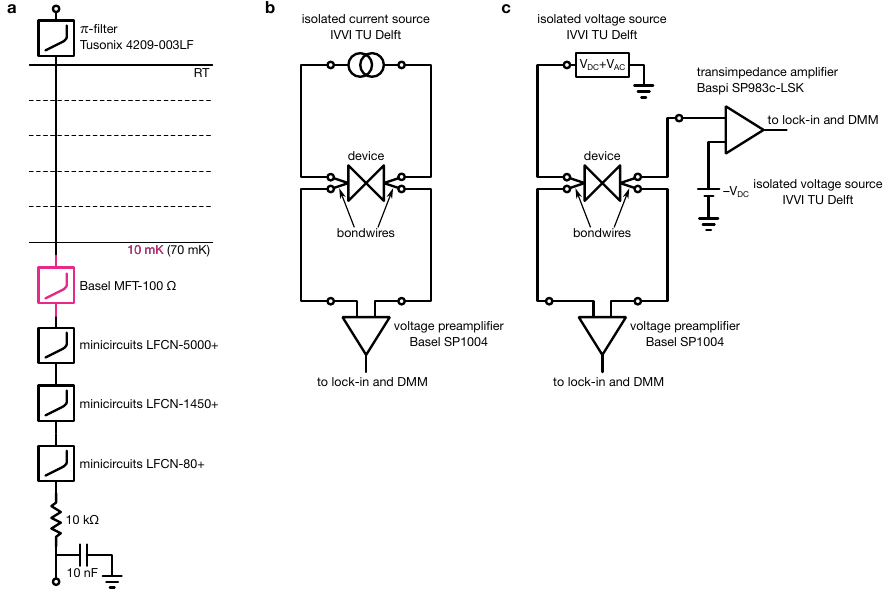}
    \caption{\textbf{Electrical measurement setup.}
    \textbf{a}, Schematic of the filtering stages used in the two dilution refrigerators.
    The measurement lines pass through a room-temperature $\pi$-filter (Tusonix 4209-003LF) and three Mini-Circuits low-pass filters (LFCN-5000+, LFCN-1450+, and LFCN-80+) at the mixing chamber, followed by an RC filter ($R=10$~\si{\kilo\ohm}, $C=10~\mathrm{nF}$) mounted on a PCB.
    In the refrigerator with a base temperature of 10~mK, an additional Basel MFT-100~$\Omega$ filter is installed at the mixing chamber; this filter and the corresponding temperature label are highlighted in magenta.
    The second refrigerator has a base temperature of 70~mK and does not contain the Basel filter.
    \textbf{b}, Current-bias configuration used for Josephson-junction measurements.
    An isolated current source (IVVI, TU Delft) biases the device, while the voltage signal is amplified by a Basel SP1004 voltage preamplifier and recorded using  a lock-in amplifier and a digital multimeter (DMM).
    \textbf{c}, Voltage-bias configuration used for QPC and SQPC measurements.
    Two isolated voltage sources (IVVI, TU Delft) provide a symmetric DC bias across the device.
    On one side, the DC bias is combined with the AC excitation and applied directly to the device, while the opposite DC bias is applied through a Basel SP983c-LSK transimpedance amplifier.
    The current signal is converted to a voltage and amplified by a Basel SP983c-LSK transimpedance amplifier, while the voltage signal across the device is amplified by a Basel SP1004 voltage preamplifier.
    The resulting signals are recorded using lock-in amplifiers and digital multimeters.}
    \slabel{fig:figS_fridge}
\end{figure*}

\begin{figure*}[h]
    \centering
    \includegraphics[width=7.09in]{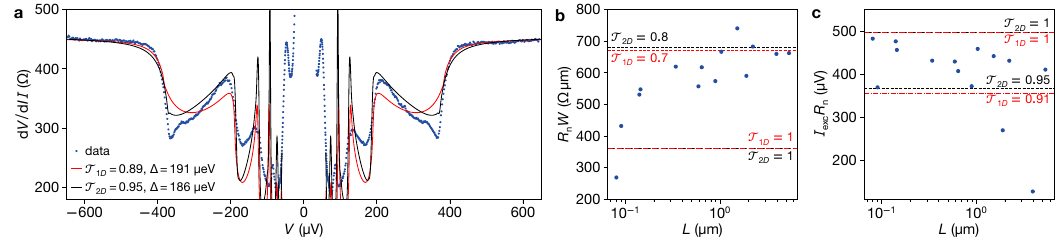}
    \caption{\textbf{Comparison of 2D and 1D models of transport in JJs.}
    The experimental data in panels \textbf{a-c} are the same as in Fig.~\textbf{2e}, the inset of Fig.~\textbf{2d}, and Fig.~\textbf{2g}, respectively.
    $\Tint_{2D}$ is the same quantity as $\Tint$ used throughout the main text -- interface transparency at normal incidence.
    Theoretical lines corresponding to $\Tint_{2D}$ take into account the distribution of incidence angle as described in Methods.
    Lines labeled as $\Tint_{1D}$, correspond to a single transparency model -- all junction modes have the same transparency, calculated using an analytical expression from Ref.~\cite{niebler2009}.}
    \slabel{fig:figS_2d_1d}
\end{figure*}

\begin{figure*}[h]
    \centering
    \includegraphics[width=7.09in]{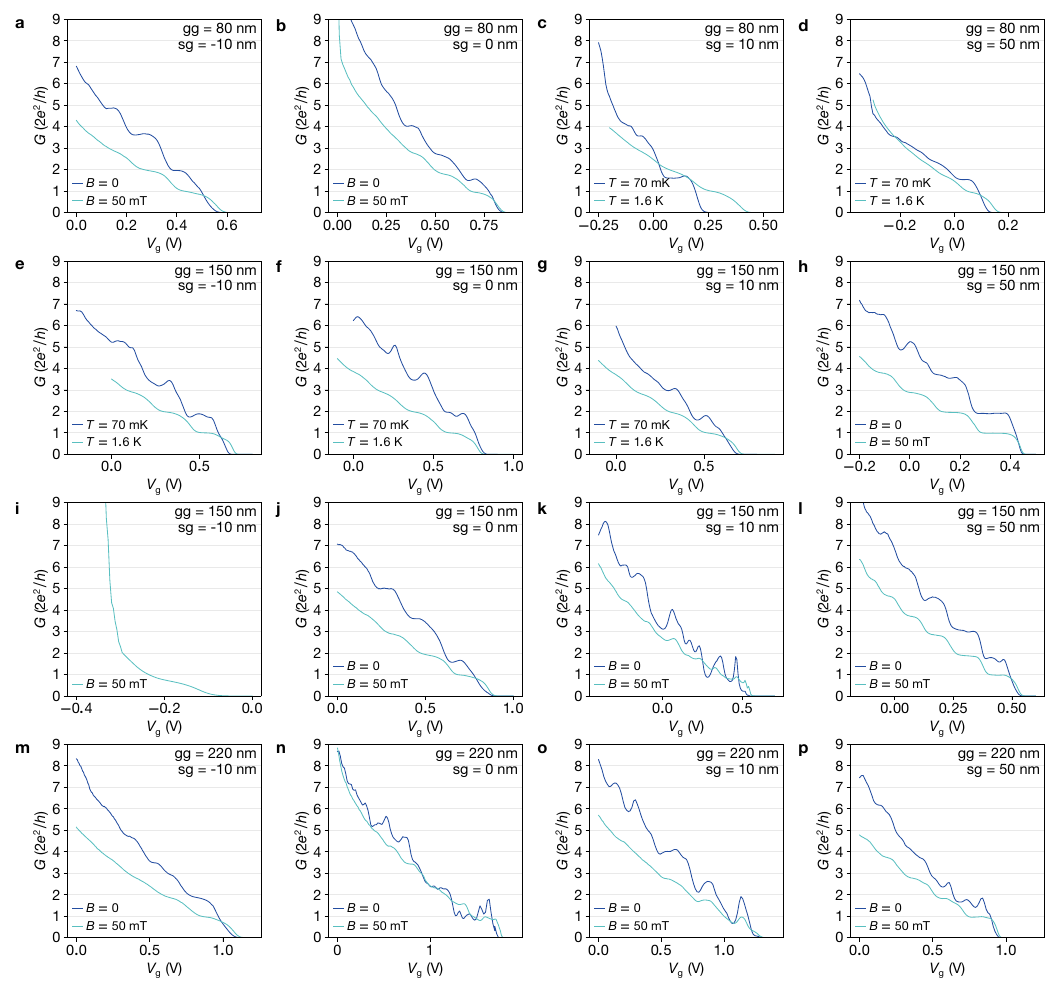}
    \caption{\textbf{Conductance vs split-gate voltage for QPC devices.}
    Conductance is measured using a quasi-four-probe configuration without subtracting any series resistance.
    Conductance in a normal state is measured either at a temperature higher than critical temperature of Al leads, or in a magnetic field perpendicular to the chip plane.
    Each panel is labeled with the nominal device geometry, where ``gg'' denotes the distance between the split gates and ``sg'' the distance between the split gates and the superconductor, with negative values indicating gate overlap with the superconductor. 
    In panel \textbf{c}, the shift between the traces is attributed to gate hysteresis, as the gates were set to zero and the sample was disconnected and reconnected between the measurements.}
    \slabel{fig:figS_steps}
\end{figure*}

\begin{figure*}[h]
    \centering
    \includegraphics[width=7.09in]{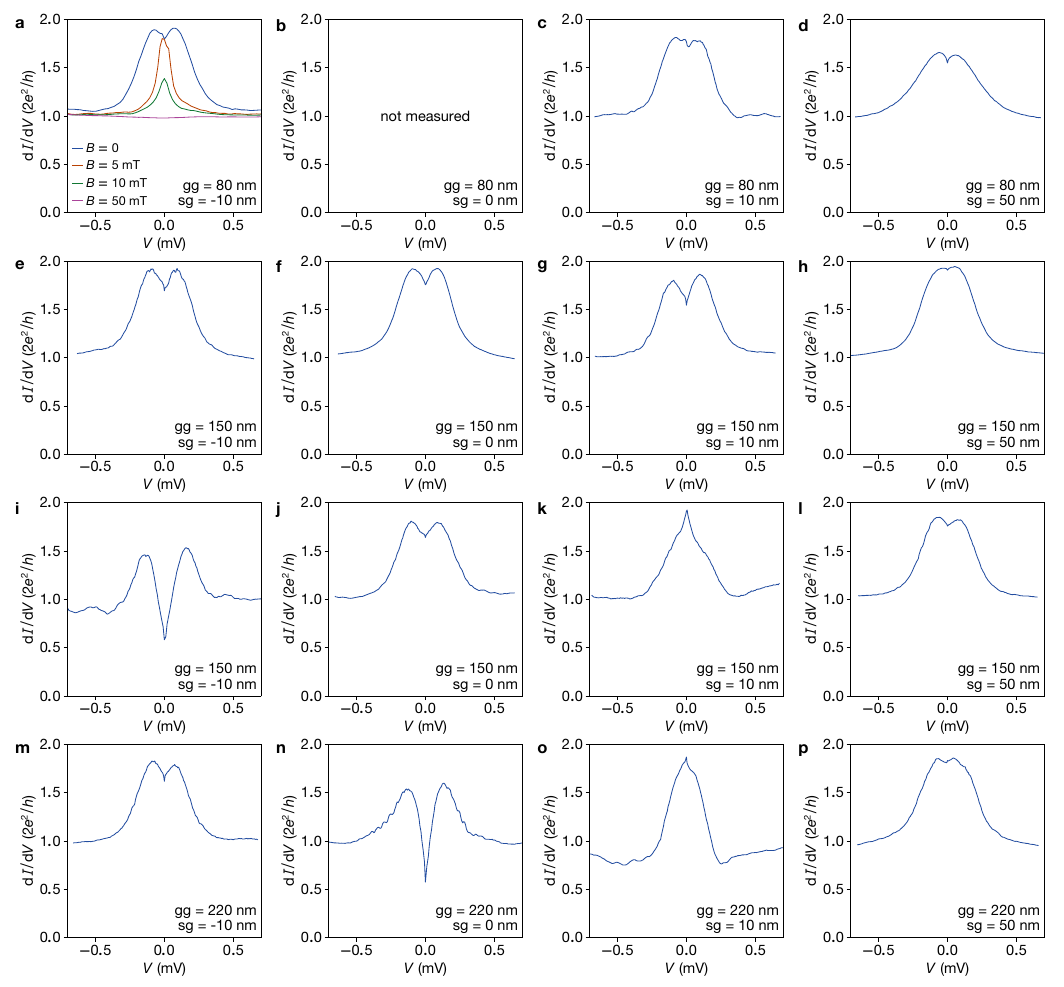}
    \caption{\textbf{Bias dependence of the differential conductance on the first plateau for QPC devices.}
    Differential conductance, $dI/dV(V)$, is measured using a quasi-four-probe configuration without subtracting any series resistance.
    The split gates are tuned such that the high-bias conductance is approximately $2e^2/h$.
    For all devices studied, the conductance is enhanced at low bias, while a dip develops near zero bias as a consequence of finite transparency.
    Each panel is labeled with the nominal device geometry, where ``gg'' denotes the distance between the split gates and ``sg'' the distance between the split gates and the superconductor, with negative values indicating gate overlap with the superconductor.
    Panel \textbf{a} illustrates the evolution of the differential conductance under a magnetic field applied perpendicular to the substrate.}
    \slabel{fig:figS_cuts_at_2G0}
\end{figure*}

\begin{figure*}[h]
    \centering
    \includegraphics[width=7.09in]{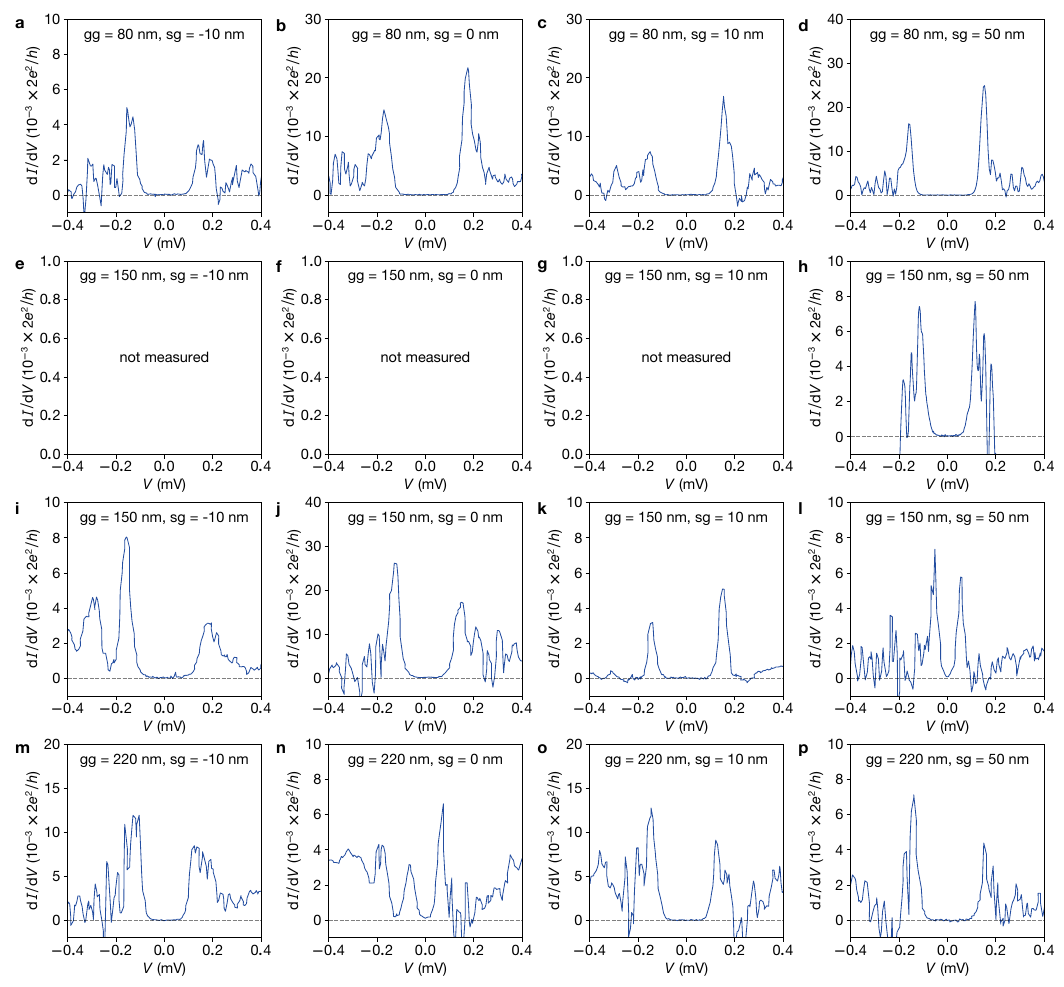}
    \caption{\textbf{Bias dependence of the differential conductance in the tunneling regime for QPC devices.}
    Differential conductance, $dI/dV(V)$, is measured using a quasi-four-probe configuration without subtracting any series resistance.
    In panel \textbf{d} one of the d.c. lines was not working and data was measured in a two-probe configuration and nominal series resistance $20.2$~k$\Omega$ was subtracted.
    The split gates are tuned such that the high-bias conductance is deep in the tunneling regime, $dI/dV \ll 2e^2/h$.
    For all devices studied, the conductance exhibits suppression at low bias, with a BCS-like peaks around $100-180$~$\mu$V.
    Each panel is labeled with the nominal device geometry, where ``gg'' denotes the distance between the split gates and ``sg'' the distance between the split gates and the superconductor, with negative values indicating gate overlap with the superconductor.}
    \slabel{fig:figS_cuts_gap}
\end{figure*}

\begin{figure*}[h]
    \centering
    \includegraphics[width=7.09in]{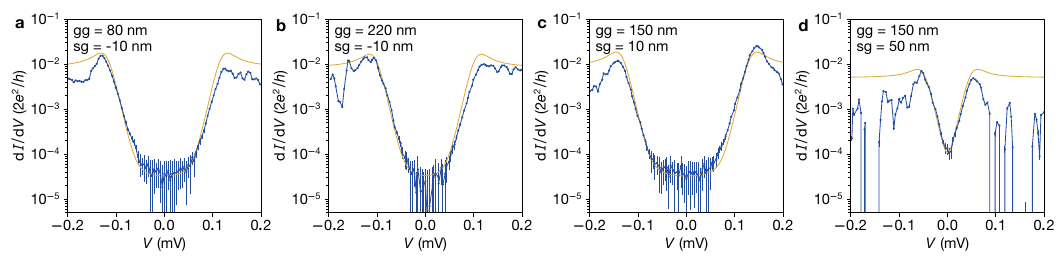}
    \caption{\textbf{Bias dependence of the differential conductance in the tunneling regime for QPC devices.}
    Differential conductance, $dI/dV(V)$, is measured using a quasi-four-probe configuration without subtracting any series resistance.
    Panels \textbf{a-d} correspond to panels of \reffig{fig:figS_cuts_gap} as \textbf{a}, \textbf{m}, \textbf{k}, \textbf{l}, respectively.
    For a better signal-to-noise ratio of low conductances at each $V$ conductance was measured 20 times with 3 time constants delay between points (lock-in filter is 3rd order, 18~dB$/$oct).
    Error bars represent the mean and standard deviation of conductance calculated from these data. 
    The orange lines show the thermally broadened BTK conductance calculated assuming a BCS-like induced gap, with $T=0.1$~K.}
    \slabel{fig:figS_cuts_gap_log_scale}
\end{figure*}

\begin{figure*}[h]
    \centering
    \includegraphics[width=6.00in]{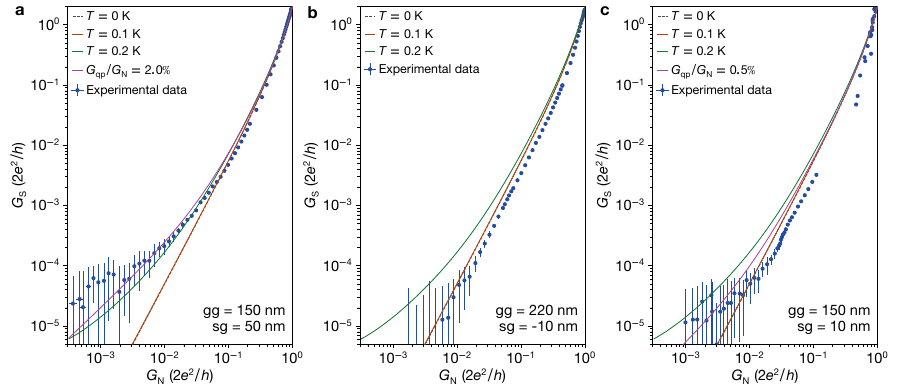}
    \caption{\textbf{Comparison of the measured superconducting conductance with the Beenakker prediction including finite-temperature and residual quasiparticle contributions.}
    Superconducting-state conductance $G_{\mathrm{S}}$ as a function of the normal-state conductance $G_{\mathrm{N}}$, when superconductivity is suppressed with magnetic field, for three QPC devices (a -- c),  here ``gg'' denotes the distance between the split gates and ``sg'' the distance between the split gates and the superconductors. 
    The $T=0$ curve corresponds to the ideal Beenakker prediction, $G_{\mathrm{S}}=2G_{\mathrm{N}}^2/(2-G_{\mathrm{N}})^2$, with conductances expressed in units of $2e^2/h$. 
    Finite-temperature curves are obtained by thermally averaging the energy-dependent BTK conductance~\cite{blonder1982}.
    To account for the excess $G_\mathrm{S}$ observed at low $G_{\mathrm{N}}$, we phenomenologically add a small residual quasiparticle density of states (DOS), such that $G_{\mathrm{S}}(T=0) = G_{\mathrm{S}}+G_{\mathrm{qp}}$,  $G_{\mathrm{qp}}=\alpha G_{\mathrm{N}}$. Here, $\alpha=G_{\mathrm{qp}}/G_{\mathrm{N}}$ parameterizes the residual quasiparticle contribution relative to the normal-state DOS. Experimental data are shown together with calculations.}
    \slabel{fig:figS_beenakker_3devices}
\end{figure*}

\begin{figure*}[h]
    \centering
    \includegraphics[width=7.09in]{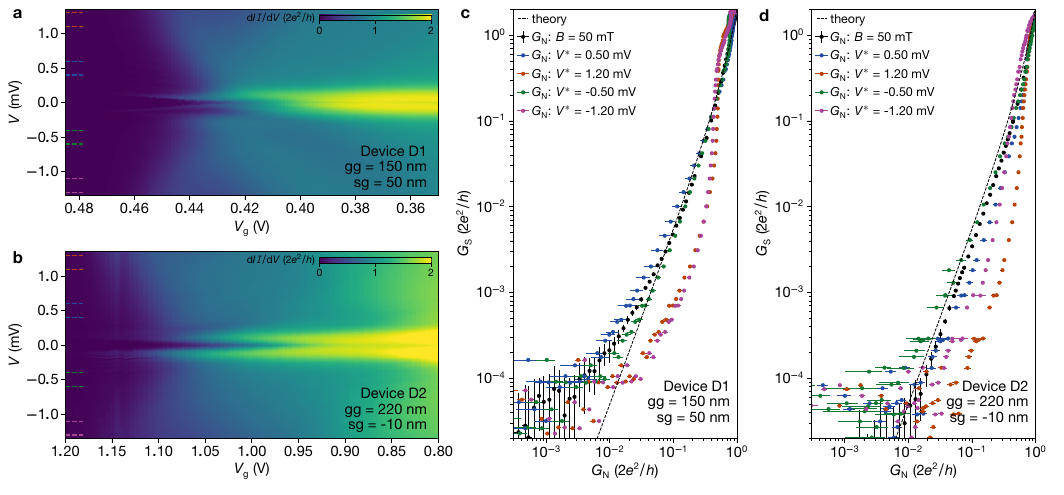}
    \caption{\textbf{Dependence of the Beenakker analysis on the extraction of the normal-state conductance.} 
    (a,b) Differential conductance as a function of bias voltage $V$ and gate voltage $V_{\mathrm{g}}$ for devices D1 and D2, respectively. 
    These devices are also presented in the main text. 
    The horizontal markers indicate the finite-bias regions used to extract the normal-state conductance $G_{\mathrm{N}}$.
    (c,d) Superconducting-state conductance $G_{\mathrm{S}}$ as a function of the corresponding normal-state conductance $G_{\mathrm{N}}$ for devices D1 and D2, respectively.
    $G_{\mathrm{S}}$ is extracted at zero bias. 
    For the finite-bias datasets, $G_{\mathrm{N}}$ is obtained by averaging the differential conductance within a bias window $V^*\pm0.1~\mathrm{mV}$ around the indicated values of $V^*$. 
    Different choices of $V^*$ therefore yield slightly different estimates of $G_{\mathrm{N}}$. 
    For comparison, the analysis using $G_{\mathrm{N}}$ measured at $B=50~\mathrm{mT}$, where superconductivity is suppressed, is also shown. 
    The dashed line denotes the ideal Beenakker prediction.}
    \slabel{fig:figS_beenakker_vg_vmeas}
\end{figure*}

\begin{figure*}[h]
    \centering
    \includegraphics[width=6.00in]{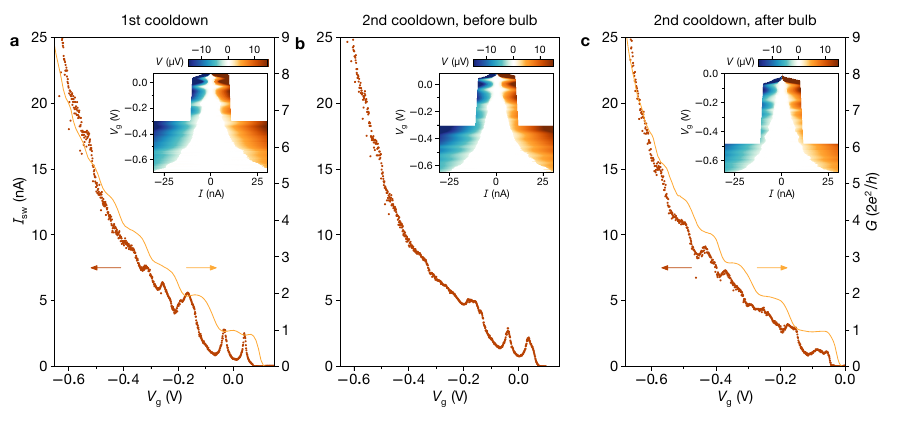}
    \caption{\textbf{Effect of bulb illumination on SQPC devices.}
    Switching current (left axis, red) and conductance (right axis, blue) of the same SQPC device ($gg=150$~\si{\nm}, $L=0.5$~\si{\um}).
    Switching current vs split gate voltage dependence of the second cooldown before bulb illumination is very close to the first cooldown.
    Bulb illumination removes sharp $\Isw$ peaks and slightly shifts the pinch-off towards more negative values.
    The $0.7$~anomaly clearly visible in the $G(\Vg)$ trace of the first cooldown is absent in the second cooldown after bulb illumination.
    Different cooldowns are complete thermal cyclings from room temperature to $70$~\si{\milli\kelvin}.}
    \slabel{fig:figS_bulb}
\end{figure*}

\begin{figure*}[h]
    \centering
    \includegraphics[width=7.09in]{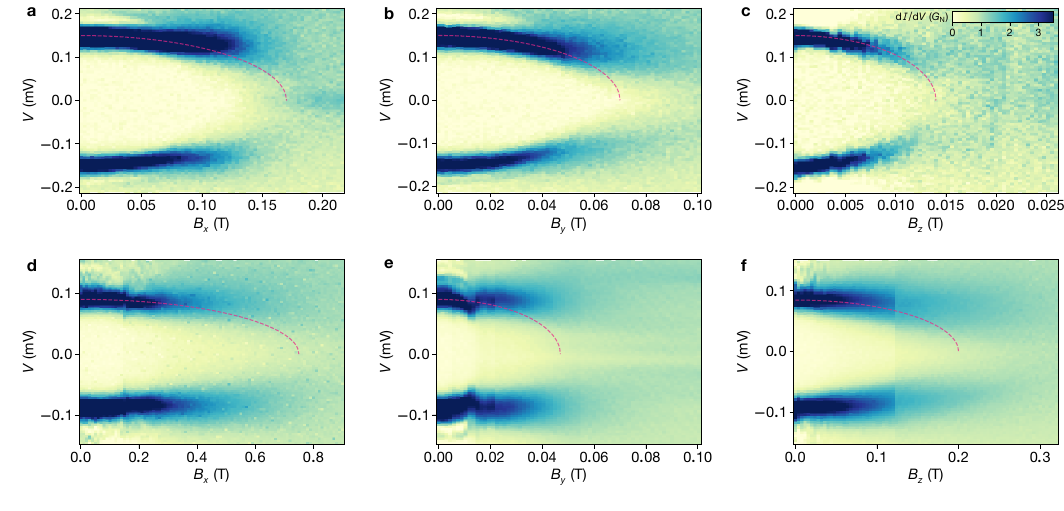}
    \caption{\textbf{Influence of contact geometry on the magnetic-field resilience of superconducting edge contacts.} Measurements on two additional devices demonstrate the reproducibility of the magnetic-field response. (a -- c) Differential conductance as a function of bias voltage and magnetic field for a first-generation etch device, with the magnetic field applied along the $x$, $z$, and $y$ directions, respectively.  (d -- f) Corresponding measurements for a deep-etch device. Here, $x$ is in-plane and parallel to the edge contact, $y$ is in-plane and perpendicular to the edge contact, and $z$ is out-of-plane. 
    Differential conductance for all panels is normalized by the high-field value, $G_\mathrm{N}$. 
    The dashed line is a guide to the eye based on the phenomenological dependence $\varepsilon_\mathrm{g}(B) \propto \sqrt{1-(B/B_\mathrm{c})^n}$ with $n=2$; note that the complex contact geometry may result in a different effective exponent.}
    \slabel{fig:figS_magnetic_field}
\end{figure*}

\begin{figure*}[h]
    \centering
    \includegraphics[width=7.09in]{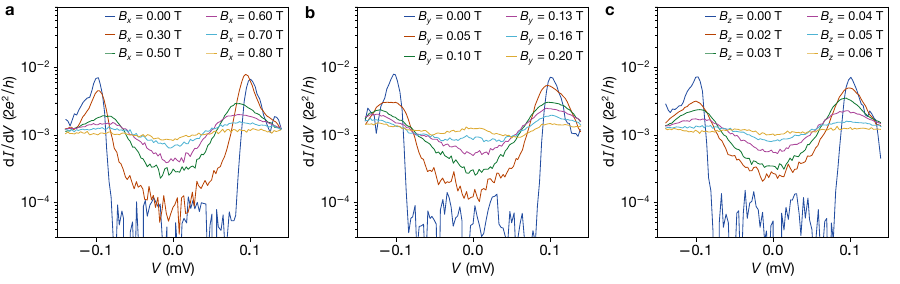}
    \caption{\textbf{Magnetic-field dependence of the superconducting gap in a deep-etch device.}
    Differential conductance as a function of bias voltage measured at selected magnetic fields for the deep-etch device presented in the main text. 
    The line cuts were measured separately from the magnetic-field maps shown in the main text and provide a more detailed view of the suppression of the superconducting gap with increasing magnetic field. 
    (a--c) Bias spectroscopy for magnetic fields applied along the $x$, $y$, and $z$ directions, respectively.
    Here, $x$ is in-plane and parallel to the edge contact, $y$ is in-plane and perpendicular to the edge contact, and $z$ is out-of-plane.}
    \slabel{fig:figS_cuts_in_magnetic_field}
\end{figure*}

\begin{figure*}[h]
    \centering
    \includegraphics[width=4.00in]{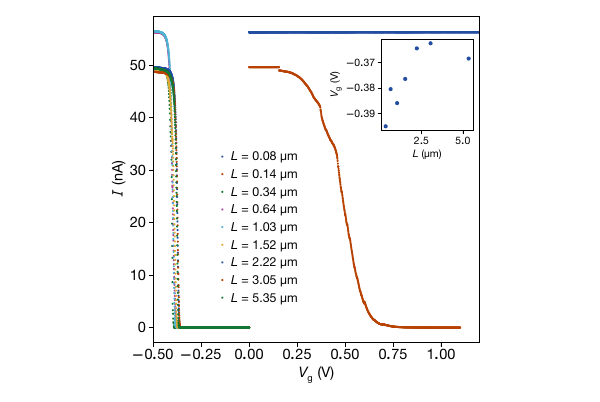}
    \caption{\textbf{Pinch-off characteristics of Josephson junction devices with different junction lengths.}
    Conductance as a function of gate voltage for JJs with different junction lengths, $L$.
    The traces correspond to the first gate sweep after cooldown, minimizing shifts of the pinch-off voltage associated with gate-voltage history.
    A voltage bias of $1~\mathrm{mV}$ was applied across the series combination of the wiring and the device.
    The inset shows the extracted pinch-off voltage as a function of junction length for devices with $L>0.3~\mu\mathrm{m}$, demonstrating comparable pinch-off voltages across this range of junction lengths.}
    \slabel{fig:figS_pinchoffs}
\end{figure*}

\begin{figure*}[h]
    \centering
    \includegraphics[width=7.09in]{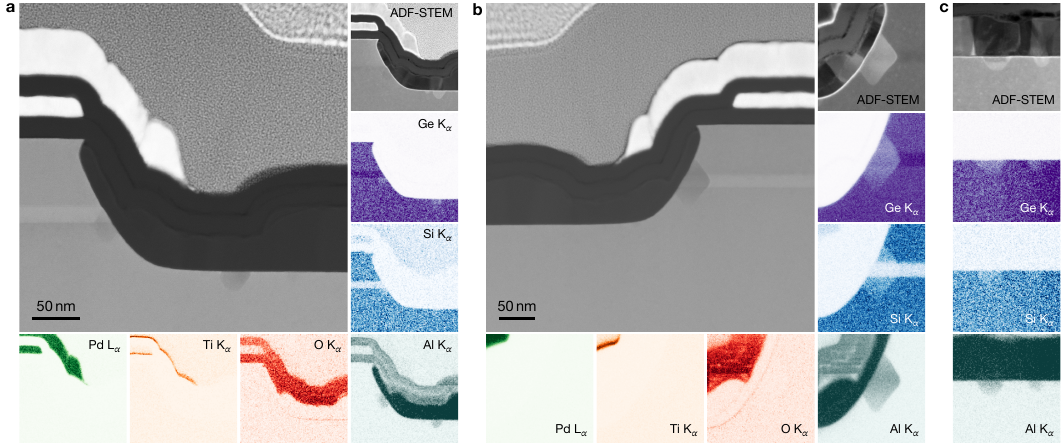}
    \caption{\textbf{Al-rich defects in Ge/SiGe heterostructures.}
    Lamellas for transmission microscopy were fabricated from QPC devices studied in this work.
    The lamella plane is not exactly in the middle of the QPC device, because in this case the split-gates would not be visible.
    \textbf{a,\,b}~HAADF-STEM micrographs of the regions with Al defects near the edge contact.
    EDS analysis reveals Al incorporation in the Ge QW and close-by SiGe spacer (\textbf{a,\,b}).
    Also, incorporation happens in the routing layer, where the QW is etched, and Al is deposited directly on SiGe (\textbf{a,\,c}).
    Similar defects were also observed in Ref.~\cite{fabris2026}.
    We attribute this incorporation to the atomic layer deposition step which is for this sample was done at $80\,^\circ$C for approximately 30~hours in total for both layers.}
    \slabel{fig:figS_EDS}
\end{figure*}

\begin{figure*}[h]
    \centering
    \includegraphics[width=7.09in]{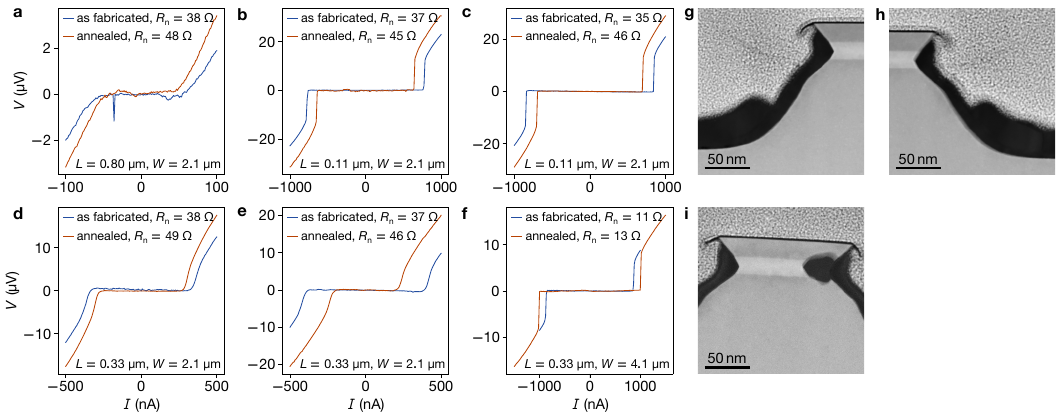}
    \caption{\textbf{Effect of annealing on Al/Ge interface.}
    Edge-contact Josephson junctions fabricated on a Ge/SiGe heterostructure that is conducting at zero gate voltage.
    This wafer has a finite carrier density without an accumulation gate, allowing us to test whether Al propagation into Ge QW is important for a good interface. 
    \textbf{a-f},~$IV$-curves of edge-contacted JJs fabricated within single fabrication layer ($T=300$~mK).
    No mesa is defined, so $\Rn$ is mostly determined by the bondpad geometry and resistivity of Ge QW.
    After contact fabrication, the sample didn't experience any high-temperature treatment like resist soft-bake or ALD, compared to a usual multilayer device (blue data).
    Afterwards, the sample was annealed in ALD for $14$~hours under similar conditions as in our $80\,^\circ$C Al$_2$O$_3$ recipe, mimicking oxide deposition but keeping precursor valves closed, and the devices were measured again (orange data).
    Normal-state resistance, $\Rn$, (see legend) was measured at $2$~K.
    \textbf{g,\,f},~HAADF-STEM micrographs of the similar pristine device.
    \textbf{i},~HAADF-STEM micrograph of one of the devices after annealing at $100\,^\circ$C.}
    \slabel{fig:figS_como_jjs}
\end{figure*}

\begin{figure*}[h]
    \centering
    \includegraphics[width=4.00in]{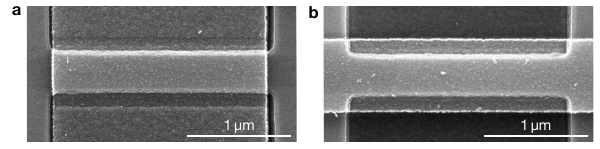}
    \caption{\textbf{JJs of different geometries.}
    SEM of two geometries of JJs studied.
    \textbf{a}~Gate climbs the mesa on top of the superconducting contact and defines the normal part of SNS junction to be of the same width as the superconductor.
    \textbf{b}~Gate covers the entire mesa region (square of \SI{40}{\um}), thus the normal region participating in transport is wider than the superconducting lead.
    Both junctions have identical geometry defined by etching and contact deposition of $W=$~\SI{2.05}{\um}, $L=$~\SI{280}{\nm}.}
    \slabel{fig:figS_SEM_JJs}
\end{figure*}

\begin{figure*}[h]
    \centering
    \includegraphics[width=5.00in]{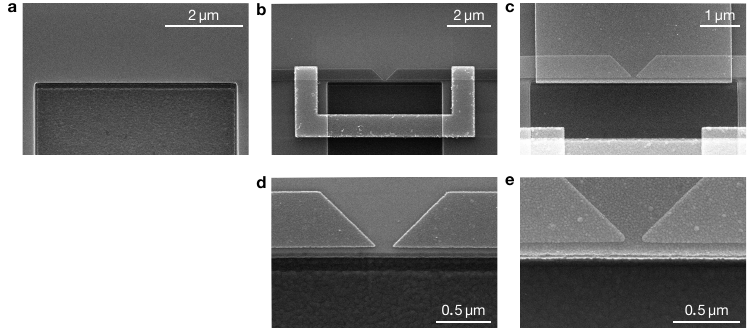}
    \caption{\textbf{Fabrication steps QPC device.}
    SEM images of the same QPC device.
    \textbf{a}~Device after etching and Al deposition.
    After \textbf{a} mesa was defined and aluminum oxide was deposited in ALD.
    \textbf{b}~First layer of gates (split gates) as well as patches were deposited.
    There were doubts that patches climb the mesa wall properly, so another patch layer was made (visible in~\textbf{c})
    After \textbf{b} one more layer of aluminum oxide was deposited, acting as a dielectric between the split gates and the global accumulation gate.
    \textbf{c}~The image is taken after deposition of the global accumulation gate (the last step of fabrication).
    \textbf{d}~and~\textbf{e} are zoomed images to the region of interest of panels~\textbf{b} and~\textbf{c} respectively.}
    \slabel{fig:figS_fab_steps}
\end{figure*}

\onecolumngrid
\clearpage
\noindent
TABLE S1. \textbf{Summary of fabricated Josephson junction devices and their transport characteristics.}
The ``GT'' column denotes the gate configuration: ``th'' indicates an accumulation gate covering the entire mesa, ``cl'' an accumulation gate climbs on top of the superconducting contacts and defines the normal part, and ``sp'' a device with split gates and a global accumulation gate.
$L_{\mathrm{SEM}}$ and $W_{\mathrm{SEM}}$ are the JJ dimensions extracted from scanning electron microscopy (SEM) images.
The ``M'' (measured) and ``W'' (working) columns indicate whether a device was measured and whether it worked, respectively.
Additional observations are listed in the ``P'' (problems) column: ``fd'' denotes a fabrication defect and ``t'' a technical problem.
$I_{\mathrm{s}}$ denotes the measured switching current in nA, $R_{\mathrm{n}}$ the normal-state resistance in $\Omega$, and $I_{\mathrm{s}}R_{\mathrm{n}}$ their product in $\mu$V.
$I_{\mathrm{exc}}$ is the excess current extracted from the high-bias current--voltage characteristic in $\mu$A.
$T$ denotes the interface transparency estimated from the excess current using the OTBK model.
Empty entries indicate that the corresponding quantity was not measured or could not be reliably extracted.
The QT5\_TR\_D2\_SQPC\_0.5 and QT5\_TR\_D1\_SQPC\_0.5 devices are presented in Fig.~\textbf{4a, b}, respectively.
\vspace{0.5em}

\label{tab:all_jjs}
\setlength{\tabcolsep}{6pt}
\csvautobooklongtable{jj.csv}
\clearpage
\twocolumngrid
\bibliography{main}